\documentclass[hyperref,namedreferences]{solarphysics}

\usepackage[optionalrh]{spr-sola-addons} 
\usepackage{graphicx}        
\usepackage{color}           
\usepackage{epstopdf}
\usepackage{tikz}

\newcommand{\etal}{{\it et al.}}
\newcommand{\eg}{{\it e.g. }}

\newcommand{\aap}{    {\it Astron. Astrophys.}}

\newcommand{\apj}{    {\it Astrophys. J.}}
\newcommand{\apjl}{   {\it Astrophys. J. Lett.}}

\newcommand{\nat}{    {\it Nature}}

\newcommand{\solphys}{{\it Solar Phys.}}
 
\newcommand{\ssr}{    {\it Space Sci. Rev.}} 
\newcommand{\apjs}{		{\it Astrophys. J. Suppl.}}
\chardef\us=`\_

\usepackage{lmodern}
\usepackage[T1]{fontenc}
\usepackage[latin9]{inputenc}
\usepackage{textcomp}
\usepackage{threeparttable}

\newcommand{\tss}[1]{\textsubscript{#1}}

\begin{document}

\begin{article}
\begin{opening}

\title{Statistical Properties of Ribbon Evolution and Reconnection Electric Fields in Eruptive and Confined Flares}

\author[addressref={aff1}, corref,email={juergen.hinterreiter@uni-graz.at}]{\inits{J.}\fnm{J.}~\lnm{Hinterreiter}}
\author[addressref={aff1,aff2},email={astrid.veronig@uni-graz.at}]{\inits{A.M.}\fnm{A.M.}~\lnm{Veronig}}
\author[addressref={aff1},email={julia.thalmann@uni-graz.at}]{\inits{J.K.}\fnm{J.K.}~\lnm{Thalmann}}
\author[addressref={aff1},email={johannes.tschernitz@edu.uni-graz.at}]{\inits{J.}\fnm{J.}~\lnm{Tschernitz}}
\author[addressref={aff2},email={werner.poetzi@uni-graz.at}]{\inits{W.}\fnm{W.}~\lnm{P{\"o}tzi}}

\address[id=aff1]{Institute of Physics/IGAM, University of Graz, Austria}
\address[id=aff2]{Kanzelh{\"o}he Observatory for Solar and Environmental Research, University of Graz, Austria}

\runningauthor{J. Hinterreiter \etal}
\runningtitle{Statistical Properties of Ribbon Evolution and Reconnetion Electric Fields}

\begin{abstract}
A statistical study of the chromospheric ribbon evolution in H$\alpha$ two-ribbon flares is performed. The data set consists of 50 confined (62\,\%) and eruptive (38\,\%) flares that occurred from June 2000 to June 2015. The flares are selected homogeneously over the H$\alpha$ and \textit{GOES} (Geostationary Operational Environmental Satellite) classes, with an emphasis on including powerful confined flares and weak eruptive flares. H$\alpha$ filtergrams from Kanzelhöhe Observatory in combination with \textit{MDI} (Michelson Doppler Imager) and \textit{HMI} (Helioseismic and Magnetic Imager) magnetograms are used to derive the ribbon separation, the ribbon-separation velocity, the magnetic-field strength, and the reconnection electric field. We find that eruptive flares reveal statistically larger ribbon separation and higher ribbon-separation velocities than confined flares. In addition, the ribbon separation of eruptive flares correlates with the GOES SXR flux, whereas no clear dependence was found for confined flares. The maximum ribbon-separation velocity is not correlated with the GOES flux, but eruptive flares reveal on average a higher ribbon-separation velocity (by $\approx$ 10\,km\,s$^{-1}$). The local reconnection electric field of confined ($cc=0.50~\pm~0.02$) and eruptive ($cc=0.77~\pm~0.03$) flares correlates with the GOES flux, indicating that more powerful flares involve stronger reconnection electric fields. In addition, eruptive flares with higher electric-field strengths tend to be accompanied by faster coronal mass ejections. 
\end{abstract}
\end{opening}

\section{Introduction}\label{sec:Introduction}

Solar flares are the most powerful eruptions on the Sun and are characterized by rapid and intense variations of the Sun's irradiance over a wide range of the electromagnetic spectrum (\eg review by \citealp{Aschwanden2005,FletcherEtAl2011}). They are powered by magnetic reconnection, during which the stored free magnetic energy in the corona is suddenly released. Solar flares present a large variety of morphological and evolutionary characteristics. They preferentially originate from complex magnetic-field configurations and may reveal complex flare-ribbon motion. In this article, we refer to flare events associated with an observed coronal mass ejection (CME) as \emph{eruptive flares} and flares that are not associated with CMEs as \emph{confined flares} \citep{Svestka1986}. The probability of flares being associated with CMEs steeply increases with the flare class. About 90 \% of X class flares are eruptive (\citealp{YashiroEtAl2006,WangZhang2007}), and all flares $\geq$ X5 tend to have an associated CME.

The most widely accepted reconnection model for eruptive flares is the so-called CSHKP model (\citealp{Carmichael1964,Sturrock1966,Hirayama1974,KoppPneuman1976}). It is intrinsically a 2.5D approach which assumes translation symmetry and successfully explains characteristic features of eruptive flares, such as quasi-parallel ribbons and their increasing separation in the course of a flare. Recently, three-dimensional numerical simulations have further increased our understanding of the physical processes involved(\eg \citealp{AulanierEtAl2012, JanvierEtAl2014}). Within the CSHKP framework, a magnetic flux system may become unstable and slowly rise to higher coronal altitudes. Below it, a current sheet develops, towards which the ambient magnetic field is drawn and forced to reconnect \citep{Vrsnak1990}. The energy released heats the local coronal plasma and accelerates particles to non-thermal energies. A significant fraction of the energy is transported towards the low solar atmosphere along newly reconnected flare loops by non-thermal electrons. They produce enhanced emission at hard X-ray (HXR) by thick-target bremsstrahlung in the low atmosphere (see \citealp{Emslie2003} and \citealp{FletcherEtAl2011}, respectively). While the HXR emission is most often observed in the form of localized kernels (``HXR footpoints''), the EUV, UV, and H$\alpha$ emission often appears in the form of elongated ribbons. They can be formed by the fast electron beams as well as by thermal conduction from the hot flaring corona.
Importantly, flare kernels and ribbons may thus be regarded as tracers of the low-atmosphere footpoints of newly reconnected coronal magnetic fields. As the reconnection region moves upwards, field lines anchored at successively larger distances from the polarity inversion line (PIL) are swept into the current sheet and reconnect. Thus, the ribbons appear further away from the PIL as the flare-loop system grows, leading to an apparent expansion motion of the H$\alpha$ flare ribbons \citep{FletcherEtAl2011}. In contrast to eruptive flares, confined flares show only a short range of separation motion of the two flare ribbons \citep{Kurokawa1989}, indicating that the reconnection region is not moving upwards.

The generation of a reconnecting current sheet is essential for the energy release in a solar flare, because the free magnetic energy stored in the corona can be dissipated and lead to particle acceleration and plasma heating \citep{MartensYoung1990, LitvinenkoSomov1995}. A general measure of the rate of magnetic reconnection is the electric voltage drop [$\dot{\varphi}$\tss{c}] along the reconnecting current sheet, which is related to the net change of magnetic flux. \citet{ForbesLin2000} showed, that the global reconnection rate can be obtained from observations as follows:

\begin{equation}
\dot{\varphi_{\mathrm{c}}}=\frac{\mathrm{d}\varphi_{\mathrm{c}}}{\mathrm{d}t}=\intop E_{\mathrm{c}}\mathrm{d}l=\frac{\partial}{\partial t}\int B_\mathrm{n}\mathrm{d}a,\label{eq:FaradaysLaw-1-1}
\end{equation}
where $E$\tss{c} is the local electric field in the coronal reconnection region, $\mathrm{d}l$ is the length along the current sheet, aligned in the direction of the ribbon, $B$\tss{n} is the component of the magnetic field normal to the photosphere, and $\mathrm{d}a$ is the newly brightened area swept by the flare ribbons. Assuming that
neither the magnetic field nor the length of the ribbons changes significantly
during a flare, one can rewrite Equation \ref{eq:FaradaysLaw-1-1} as follows (see \citealp{ForbesLin2000} and references therein):

\begin{equation}
\dot{\varphi_{\mathrm{c}}}=\int v_{\mathrm{r}}B_{\mathrm{n}}\mathrm{d}l,\label{eq:FaradaysLaw-1-1-1}
\end{equation}
where $v$\tss{r} is the ribbon-separation velocity. \citet*{QiuEtAl2002} pointed out that for a two-ribbon flare with a 2D configuration (\textit{i.e.} translation symmetry along the ribbon) and the line-tying nature of the photospheric magnetic field, Equation~\ref{eq:FaradaysLaw-1-1} and Equation \ref{eq:FaradaysLaw-1-1-1} reduces to (see also \citealp{ForbesPriest1984,ForbesLin2000}):

\begin{equation}
E_{\mathrm{c}}=v_{\mathrm{r}}B_{\mathrm{n}},\label{eq:E=00003DvxB-1}
\end{equation}
where $E$\tss{c} can be interpreted as a local reconnection rate. 

When applying Equation \ref{eq:E=00003DvxB-1}, the outer front of the flare ribbons should be considered, because this part is related to the newly reconnected field lines along which the accelerated particles travel downwards to the solar surface. Since the flare ribbons are tracked using chromospheric images, the chromospheric magnetic field should also be used to determine the reconnection electric field using Equation \ref{eq:E=00003DvxB-1}. In practice, however, the chromospheric magnetic field is difficult to measure, so that generally photospheric magnetic-field maps are used to retrieve the reconnection rates. 

Equation \ref{eq:E=00003DvxB-1} was applied in various case studies of solar flares \citep{PolettoKopp1986, QiuEtAl2002,AsaiEtAl2004,TemmerEtAl2007, MiklenicEtAl2007}. \citet{LiuWang2009} and \citet{JingEtAl2005} each performed statistical studies of powerful and mainly eruptive flares. In both studies, the authors found a clear dependence of the local coronal electric field on the strength of the flare as indicated by the soft X-ray (SXR) peak flux measured by \textit{GOES} (Geostationary Operational Environmental Satellite).

In this article we present the first systematic statistical study comparing the electric field in the reconnecting current sheet in eruptive and confined flares using a homogeneous data set that spans more than a solar cycle. The set covers in total 50 events ranging from GOES classes B to > X10, including 19 eruptive and 31 confined flares.

\section{Data and Data Reduction}

\begin{figure}
\centerline{\includegraphics[width=0.5\textwidth,clip=]{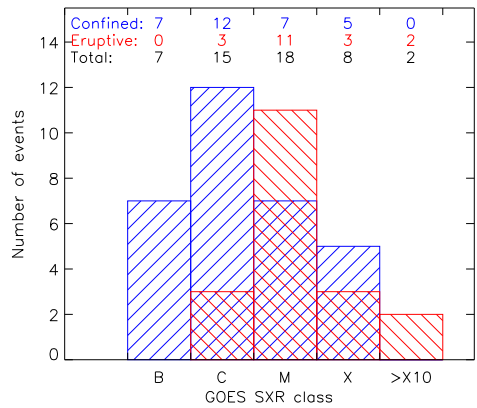}}
              \caption{Distribution of the selected flares (50 in total, 19 eruptive and 31 confined). Blue and red bars correspond to eruptive and confined flares, respectively.}
\label{HistoSelectedFlares}
\end{figure}

The data set consists of 50 H$\alpha$ flares, selected to contain all powerful flares and an appropriate number of weaker flares (eruptive and confined) that originated from close to the central meridian between June 2000 and June 2015 and that were observed in full-disk H$\alpha$ filtergrams at Kanzelhöhe Observatory for Solar and Environmental Research (KSO\footnote{www.kso.ac.at}).

We aimed at having a good coverage over H$\alpha$ and GOES classes with a balance between confined and eruptive flares. Figure \ref{HistoSelectedFlares} shows the distribution of the selected flares over the GOES class. First we searched for all flares of H$\alpha$ classification\footnote{sidc.be/educational/classification.php\#OClass} 4 and 3. Then an appropriate number of importance class 2 flares were selected. For the importance class 1 and S flares we were looking for suitable flares beginning from 2015 and going backwards in time. In addition, we put an emphasis on including powerful confined as well as weak eruptive flares. The flares are selected to be close to the center of the solar disk (CMD < 45\textdegree{}), in order to minimize projection effects. The central meridian distance (CMD) is the angular distance in solar longitude measured from the central meridian. 

The {\textit{SOHO/LASCO} (Solar and Heliospheric Observatory/Large Angle and Spectrometric Coronagraph Experiment) CME catalog}\footnote{cdaw.gsfc.nasa.gov/CME\_list} was checked, to determine the flare--CME association. The flare position had to be consistent with the position angle given in the CME catalog and the flare had to occur within 60 minutes of the linearly extrapolated starting time of the CME. For the M1.2/1N flare on 1 October 2011 an eruption in the original LASCO movie can be seen, but no entry in the SOHO/LASCO CME catalog exists. Therefore, we refer to \cite{TemmerEtAl2017}, who report a CME speed of 450\,km\,s$^{-1}$.

To track the flare-ribbon-separation motion, we used H$\alpha$ full-disk data obtained at the KSO. The KSO H$\alpha$ telescope is a refractor with an aperture ratio of d/f=100/2000 equipped with a Lyot filter centered at 6563\,\textrm{Å} and a FWHM of 0.7\,\textrm{Å}. For the time range June 2000 to April 2008 the resolution of the images was about 2.2$''$ (8-bit CCD until mid 2005 and 10-bit CCD until 2008) with a temporal cadence of roughly one minute. Since April 2008, KSO has obtained high-resolution (approximately 1$''$, 12-bit CCD) and high-cadence (roughly six seconds) filtergrams. In addition, all images have to pass a primary quality check \citep{PoetziEtAl2015}. In order to include also the most powerful flares during this time range, we also used H$\alpha$ data from other observatories. To analyze the X17.2/4B flare on 28 October 2003, we used high-resolution H$\alpha$ filtergrams obtained by Udaipur Solar Observatory (USO) with a 15-cm aperture f/15 telescope and a 12-bit CCD. The temporal cadence of the images is approximately 30 seconds and the pixel scale was derived by co-alignment with KSO data, which partially covered the event, resulting in 0.6 arcsecs. The images for the X10.0/2B flare on 29 October 2003 are provided by the National Solar Observatory at Sacramento Peak. They were obtained by a 12-bit CCD camera with a pixel size of about one arcsec and a temporal cadence of about one minute \citep{NeidigEtAl1998}. 

To calculate the coronal electric field, measurements of the photospheric magnetic
field are required. Therefore, we used 96m full-disk magnetograms from {\it Michelson Doppler Imager} (MDI; \citealp{ScherrerEtAl1995}) onboard SOHO for flares before 2011 and low-noise 720-second magnetograms from {\it Helioseismic and Magnetic Imager} onboard {\it Solar Dynamics Observatory} (SDO/HMI; \citealp{SchouEtAl2012}) for flares since 2011. For each event we selected the latest available magnetogram before the flare start. 

\begin{table*}
\caption[Results for the selected flares]{Results of ribbon tracking for all flares under study. Given are the date, KSO flare times, flare classification (H$\alpha$ and GOES), heliographic position ($u$ is the angular distance to the solar disc center) and CME properties. The CME speed is the linear speed from LASCO catalog. The ribbon distance is the minimum and maximum distance of the flare ribbons. Ribbon-separation velocity gives the maximum speed of the faster ribbon. $B_{\mathrm{E}}$ is the magnetic field at the leading front of the flare ribbon at the time of the maximum electric field. $E_{\mathrm{c}}$ is the maximum electric field.}
\begin{threeparttable} {}
\setlength\tabcolsep{2pt}{\tiny \par}
\begin{tiny}
\begin{tabular}{c|c|c|c|c|c|c|c|c|r@{\extracolsep{0pt}.}l|r@{\extracolsep{0pt}.}l|r@{\extracolsep{0pt}.}l|r@{\extracolsep{0pt}.}l|r@{\extracolsep{0pt}.}l}
\multicolumn{1}{c}{} & \multicolumn{3}{c}{} & \multicolumn{2}{c}{} & \multicolumn{2}{c}{} & \multicolumn{1}{c}{} & \multicolumn{4}{c}{} & \multicolumn{2}{c}{} & \multicolumn{2}{c}{} & \multicolumn{2}{c}{}\tabularnewline
Date & \multicolumn{3}{c|}{KSO times} & \multicolumn{2}{c|}{Class} & \multicolumn{2}{c|}{Position} & \multicolumn{1}{c|}{CME} & \multicolumn{4}{c|}{Ribbon distance} & \multicolumn{2}{c|}{Velocity} & \multicolumn{2}{c|}{$B_{\mathrm{E}}$} & \multicolumn{2}{c}{$E_{\mathrm{c}}$}\tabularnewline
\hline 
 & Start & Max & End & H$\alpha$ & GOES & Lat.Lon & $u$ & Speed & \multicolumn{2}{c|}{Min} & \multicolumn{2}{c|}{Max} & \multicolumn{2}{c|}{Max} & \multicolumn{2}{c|}{} & \multicolumn{2}{c}{Max}\tabularnewline
YYYY Mon DD & UT & UT & UT &  &  &  & {[}\textdegree {]} & {[}\,km\,s$^{-1}${]} & \multicolumn{2}{c|}{{[}Mm{]}} & \multicolumn{2}{c|}{{[}Mm{]}} & \multicolumn{2}{c|}{{[}\,km\,s$^{-1}${]}} & \multicolumn{2}{c|}{{[}G{]}} & \multicolumn{2}{c}{{[}\,V\,cm$^{-1}${]}}\tabularnewline
\hline 
2000 Jun. 01 & 07:30 & 07:32 & 08:18 & 2N & C8.2 & S13E24 & 27.4 & -  & 11&8\textpm{}4.0 & 38&4\textpm{}4.3 & 23&9\textpm{}10.6 & \multicolumn{2}{c|}{623} & 10&6\textpm{}3.1\tabularnewline
2000 Jul. 19{*} & 06:37 & 07:23 & 09:01 & 3N & M6.4 & S15E07 & 12.3 & -  & 15&3\textpm{}7.0 & 42&4\textpm{}4.6 & 13&0\textpm{}8.4 & \multicolumn{2}{c|}{1618} & 15&6\textpm{}13.9\tabularnewline
2000 Sep. 12 & 11:22 & 12:00 & 14:58 & 2F & M1.0 & S19W08 & 14.1 & 1550 & 51&5\textpm{}5.6 & 73&7\textpm{}4.6 & 10&4\textpm{}6.9 & \multicolumn{2}{c|}{231} & 2&3\textpm{}1.1\tabularnewline
2001 Aug. 25 & 16:23 & 16:32 & 17:25 & 3N & X5.3 & S21E38 & 39.3 & 1433 & 48&3\textpm{}6.6 & 83&9\textpm{}5.6 & 23&7\textpm{}12.1 & \multicolumn{2}{c|}{1318} & 28&4\textpm{}6.9\tabularnewline
2003 Oct. 26{*} & 06:46 & 06:46 & 09:17 & 3B & X1.2 & S14E41 & 41.4 & 1371 & 35&0\textpm{}3.9 & 46&2\textpm{}4.0 & 7&3\textpm{}8.8 & \multicolumn{2}{c|}{1497} & 6&0\textpm{}8.1\tabularnewline
2003 Oct. 28 & 10:32 & 11:23 & 14:20 & 4B & X17.2 & S16E07 & 13.2 & 2459 & 14&7\textpm{}0.9 & 65&4\textpm{}1.0 & 56&5\textpm{}4.2 & \multicolumn{2}{c|}{1693} & 68&1\textpm{}3.4\tabularnewline
2003 Oct. 29 & 20:37 & 20:42 & 22:53 & 2B & X10.0 & S15W02 & 10.6 & 2029 & 21&5\textpm{}2.4 & 54&8\textpm{}3.9 & 25&2\textpm{}5.8 & \multicolumn{2}{c|}{2389} & 60&3\textpm{}7.2\tabularnewline
2003 Nov. 18 & 07:30 & 07:50 & 11:04 & 3N & M3.2 & S02E37 & 37.0 & 1660 & 42&7\textpm{}4.9 & 63&6\textpm{}4.8 & 45&8\textpm{}9.1 & \multicolumn{2}{c|}{683} & 21&9\textpm{}2.8\tabularnewline
2003 Nov. 20 & 07:35 & 07:42 & 08:43 & 3B & M9.6 & N01W08 & 8.6 & 669 & 18&8\textpm{}5.8 & 73&3\textpm{}6.4 & 38&0\textpm{}14.6 & \multicolumn{2}{c|}{412} & 13&3\textpm{}3.1\tabularnewline
2004 Jul. 16 & 13:50 & 13:57 & 14:31 & 4B & X3.6 & S09E29 & 29.1 &  - & 58&7\textpm{}3.3 & 78&3\textpm{}3.3 & 29&4\textpm{}11.5 & \multicolumn{2}{c|}{565} & 10&5\textpm{}5.7\tabularnewline
2004 Jul. 17 & 07:54 & 08:05 & 08:53 & 2F & X1.0 & S11E22 & 22.7 &  - & 50&8\textpm{}3.3 & 56&3\textpm{}3.3 & 12&1\textpm{}14.1 & \multicolumn{2}{c|}{585} & 6&8\textpm{}4.1\tabularnewline
2004 Jul. 20 & 12:26 & 12:31 & 13:30 & 3B & M8.6 & N10E32 & 35.2 & 710 & 22&7\textpm{}3.3 & 53&5\textpm{}3.3 & 37&6\textpm{}15.1 & \multicolumn{2}{c|}{632} & 19&4\textpm{}4.7\tabularnewline
2005 Jan. 15 & 11:46 & 11:51 & 12:00 & 2F & M1.2 & N13E01 & 8.4 &  - & 34&4\textpm{}5.8 & 46&0\textpm{}5.2 & 20&4\textpm{}13.7 & \multicolumn{2}{c|}{2418} & 19&0\textpm{}15.8\tabularnewline
2005 Jan. 17 & 07:16 & 09:51 & 11:57 & 3B & X3.8 & N14W24 & 25.4 & 2547 & 17&5\textpm{}8.5 & 70&1\textpm{}5.8 & 46&4\textpm{}9.1 & \multicolumn{2}{c|}{1360} & 42&7\textpm{}11.3\tabularnewline
2005 May 12 & 07:28 & 07:34 & 08:57 & 2B & M1.6 & N12E28 & 29.2 &  - & 53&7\textpm{}4.5 & 68&4\textpm{}4.4 & 39&0\textpm{}14.8 & \multicolumn{2}{c|}{207} & 3&0\textpm{}1.6\tabularnewline
2005 Sep. 12 & 08:42 & 08:49 & 11:05 & 3N & M6.1 & S13E25 & 25.3 &  - & 14&0\textpm{}3.9 & 42&7\textpm{}6.2 & 39&5\textpm{}12.1 & \multicolumn{2}{c|}{204} & 8&0\textpm{}2.5\tabularnewline
2005 Sep. 15{*} & 08:34 & 08:40 & 10:10 & 2N & X1.1 & S11W15 & 15.3 &  - & 38&7\textpm{}5.2 & 42&7\textpm{}4.0 & 4&2\textpm{}12.4 & \multicolumn{2}{c|}{805} & 3&1\textpm{}4.6\tabularnewline
2006 Jul. 06 & 08:16 & 08:42 & 10:24 & 3N & M2.5 & S10W30 & 30.5 & 911 & 40&0\textpm{}4.9 & 71&8\textpm{}5.2 & 62&9\textpm{}22.5 & \multicolumn{2}{c|}{683} & 42&9\textpm{}8.3\tabularnewline
2011 Mar. 07 & 13:48 & 14:31 & 14:50 & 2F & M1.9 & N10E18 & 18.0 & 698 & 30&9\textpm{}1.6 & 50&7\textpm{}2.1 & 14&5\textpm{}7.1 & \multicolumn{2}{c|}{160} & 2&0\textpm{}0.6\tabularnewline
2011 Apr. 22{*} & 11:09 & 11:33 & 12:02 & 2N & C7.7 & S16E34 & 39.7 &  - & 25&1\textpm{}2.1 & 31&6\textpm{}1.5 & 7&5\textpm{}7.5 & \multicolumn{2}{c|}{761} & 5&7\textpm{}2.9\tabularnewline
2011 Jun. 02 & 07:25 & 07:47 & 08:11 & 2N & C3.7 & S19E20 & 27.7 & 976 & 23&9\textpm{}1.9 & 37&3\textpm{}2.7 & 9&1\textpm{}8.8 & \multicolumn{2}{c|}{403} & 2&1\textpm{}1.1\tabularnewline
2011 Sep. 28 & 12:29 & 12:34 & 12:55 & 1N & C9.3 & N15W01 & 21.9 &  - & 38&0\textpm{}1.5 & 46&6\textpm{}1.5 & 26&1\textpm{}6.3 & \multicolumn{2}{c|}{92} & 2&0\textpm{}3.6\tabularnewline
2011 Oct. 01 & 09:23 & 10:00 & 10:38 & 1N & M1.2 & N08W03 & 15.0 & 450 & 16&6\textpm{}2.9 & 39&0\textpm{}1.5 & 29&2\textpm{}6.7 & \multicolumn{2}{c|}{1322} & 22&9\textpm{}3.8\tabularnewline
2011 Nov. 09 & 13:06 & 13:27 & 14:15 & 2N & M1.1 & N22E36 & 43.5 & 907 & 27&9\textpm{}1.5 & 68&1\textpm{}1.5 & 59&2\textpm{}6.1 & \multicolumn{2}{c|}{134} & 5&9\textpm{}1.6\tabularnewline
2012 Mar. 06 & 12:23 & 12:40 & 13:26 & 2N & M2.1 & N17E35 & 35.5 &  - & 11&2\textpm{}2.3 & 17&2\textpm{}2.2 & 17&1\textpm{}5.9 & \multicolumn{2}{c|}{822} & 12&0\textpm{}2.8\tabularnewline
2012 Mar. 15 & 07:25 & 07:45 & 08:45 & 2F & M1.8 & N14E00 & 6.8 & 485 & 28&2\textpm{}2.1 & 47&5\textpm{}2.7 & 22&7\textpm{}8.3 & \multicolumn{2}{c|}{589} & 11&3\textpm{}1.8\tabularnewline
2012 Apr. 27 & 08:11 & 08:21 & 08:42 & 1N & M1.0 & N12W30 & 30.6 &  - & 19&6\textpm{}1.5 & 28&2\textpm{}2.4 & 17&6\textpm{}6.4 & \multicolumn{2}{c|}{572} & 1&3\textpm{}1.4\tabularnewline
2012 Jul. 10 & 06:10 & 06:23 & 07:34 & 1F & M2.1 & S16E30 & 31.9 &  - & 8&6\textpm{}5.2 & 12&7\textpm{}1.5 & 3&1\textpm{}7.0 & \multicolumn{2}{c|}{873} & 2&7\textpm{}3.1\tabularnewline
2013 Apr. 11 & 06:56 & 07:08 & 09:15 & 3B & M6.5 & N08E14 & 14.1 & 861 & 18&0\textpm{}2.1 & 47&0\textpm{}1.9 & 39&1\textpm{}5.2 & \multicolumn{2}{c|}{354} & 10&4\textpm{}1.7\tabularnewline
2013 Jul. 09 & 13:27 & 13:32 & 13:48 & SN & C2.3 & S10W21 & 21.7 &  - & 35&0\textpm{}1.5 & 42&2\textpm{}1.5 & 4&4\textpm{}5.6 & \multicolumn{2}{c|}{613} & 1&7\textpm{}2.1\tabularnewline
2013 Aug. 02 & 11:10 & 11:11 & 11:24 & SF & B9.7 & S15W10 & 13.4 &  - & 13&0\textpm{}2.3 & 16&5\textpm{}2.5 & 5&0\textpm{}7.2 & \multicolumn{2}{c|}{125} & 0&5\textpm{}0.4\tabularnewline
2013 Aug. 11 & 12:29 & 12:31 & 12:42 & SF & B7.1 & S21E31 & 33.3 &  - & 24&9\textpm{}2.0 & 28&9\textpm{}8.2 & 8&0\textpm{}11.7 & \multicolumn{2}{c|}{418} & 3&3\textpm{}2.5\tabularnewline
2013 Sep. 23{*} & 07:10 & 07:11 & 07:24 & SF & B6.0 & N10E35 & 38.8 &  - & 27&9\textpm{}1.5 & 31&5\textpm{}7.8 & 17&6\textpm{}9.1 & \multicolumn{2}{c|}{104} & 1&8\textpm{}0.7\tabularnewline
2013 Oct. 16 & 09:12 & 09:20 & 09:44 & SF & C1.9 & S09W42 & 41.8 &  - & 14&3\textpm{}2.1 & 18&2\textpm{}2.1 & 2&1\textpm{}4.4 & \multicolumn{2}{c|}{305} & 0&6\textpm{}0.7\tabularnewline
2013 Oct. 20 & 08:36 & 08:41 & 09:08 & 1N & C2.9 & N22W32 & 41.7 & 398 & 17&0\textpm{}1.6 & 31&9\textpm{}1.6 & 35&8\textpm{}10.2 & \multicolumn{2}{c|}{226} & 5&1\textpm{}1.3\tabularnewline
2013 Nov. 29 & 09:55 & 10:10 & 10:14 & 1F & C1.5 & S06W23 & 23.5 &  - & 12&4\textpm{}6.8 & 18&1\textpm{}2.8 & 7&6\textpm{}8.3 & \multicolumn{2}{c|}{35} & 0&1\textpm{}0.1\tabularnewline
2013 Dec. 14 & 11:06 & 11:19 & 11:58 & 1F & C2.3 & S14W14 & 20.3 &  - & 37&9\textpm{}1.5 & 44&6\textpm{}1.5 & 4&7\textpm{}7.1 & \multicolumn{2}{c|}{332} & 1&4\textpm{}1.1\tabularnewline
2013 Dec. 28 & 12:42 & 12:44 & 13:05 & 1F & C3.0 & S17E10 & 21.9 &  - & 16&7\textpm{}1.9 & 23&5\textpm{}3.1 & 12&2\textpm{}6.7 & \multicolumn{2}{c|}{316} & 3&6\textpm{}1.5\tabularnewline
2014 Feb. 14 & 10:38 & 10:40 & 11:04 & 1N & C7.2 & S11W29 & 33.9 &  - & 28&2\textpm{}1.5 & 37&1\textpm{}6.2 & 11&6\textpm{}5.7 & \multicolumn{2}{c|}{1295} & 13&6\textpm{}3.7\tabularnewline
2014 Mar. 21 & 10:18 & 10:35 & 11:01 & 1F & C2.7 & N17E39 & 39.3 & 423 & 28&6\textpm{}5.3 & 42&4\textpm{}2.8 & 5&7\textpm{}4.7 & \multicolumn{2}{c|}{436} & 1&8\textpm{}1.2\tabularnewline
2014 May 02 & 09:17 & 09:23 & 10:19 & 1N & C4.4 & S19W16 & 27.9 &  - & 20&6\textpm{}1.5 & 34&4\textpm{}1.6 & 14&7\textpm{}7.5 & \multicolumn{2}{c|}{1282} & 8&4\textpm{}4.8\tabularnewline
2014 May 10 & 06:51 & 07:01 & 08:02 & 2N & C8.7 & N03E27 & 27.0 &  - & 13&9\textpm{}3.0 & 39&8\textpm{}2.8 & 32&5\textpm{}5.2 & \multicolumn{2}{c|}{836} & 22&5\textpm{}2.4\tabularnewline
2014 May 12 & 06:25 & 06:38 & 07:07 & 1F & C2.3 & N04W02 & 2.2 &  - & 15&5\textpm{}4.0 & 31&2\textpm{}1.5 & 26&7\textpm{}6.5 & \multicolumn{2}{c|}{118} & 2&1\textpm{}0.5\tabularnewline
2014 Jun. 21 & 13:36 & 13:54 & 14:03 & SF & B4.7 & S11E04 & 10.1 &  - & 32&8\textpm{}1.6 & 40&7\textpm{}2.8 & 8&4\textpm{}3.8 & \multicolumn{2}{c|}{675} & 5&5\textpm{}1.9\tabularnewline
2014 Jun 26 & 07:12 & 07:31 & 07:46 & SN & B3.1 & N10E30 & 32.3 &  - & 25&0\textpm{}1.9 & 32&1\textpm{}1.5 & 7&6\textpm{}9.7 & \multicolumn{2}{c|}{66} & 0&4\textpm{}0.3\tabularnewline
2014 Aug. 10 & 10:05 & 10:07 & 10:14 & SF & B8.9 & S21W12 & 18.7 &  - & 23&8\textpm{}4.6 & 34&5\textpm{}2.5 & 26&9\textpm{}19.2 & \multicolumn{2}{c|}{545} & 7&9\textpm{}2.6\tabularnewline
2014 Oct. 22{*} & 14:02 & 14:06 & 14:55 & 3B & X1.6 & S14E15 & 17.2 &  - & 47&0\textpm{}2.6 & 52&6\textpm{}2.0 & 6&1\textpm{}7.6 & \multicolumn{2}{c|}{991} & 6&0\textpm{}3.8\tabularnewline
2014 Oct. 26{*} & 10:03 & 10:51 & 10:51 & 2N & X2.0 & S14W34 & 34.7 &  - & 27&8\textpm{}1.5 & 29&2\textpm{}1.7 & 3&1\textpm{}6.7 & \multicolumn{2}{c|}{1750} & 5&5\textpm{}5.9\tabularnewline
2014 Nov. 02 & 13:07 & 13:11 & 13:19 & SF & B7.6 & S04E29 & 28.9 &  - & 16&0\textpm{}8.2 & 22&5\textpm{}4.6 & 6&6\textpm{}4.4 & \multicolumn{2}{c|}{393} & 1&5\textpm{}2.2\tabularnewline
2015 Jun. 25{*} & 08:02 & 08:14 & 12:00 & 3B & M7.9 & N11W41 & 42.9 & 1627 & 21&9\textpm{}1.9 & 40&1\textpm{}2.6 & 10&3\textpm{}10.0 & \multicolumn{2}{c|}{1424} & 10&7\textpm{}0.9\tabularnewline
\hline
\end{tabular}{\tiny \par}
\end{tiny}
		\begin{tablenotes}
        \item[*] Tracking of the two flare ribbons is done separately, because the ribbons do not lie \textit{vis-à-vis} to each other (see Figure \ref{20141022FlareEvolution} for an example).
    \end{tablenotes}
	\end{threeparttable}
\normalsize
\label{tab:ResultsTable}
\end{table*}

Furthermore, the GOES SXR light curves in the $1-8$\,\textrm{Å} band were used to quantify the flare energy release. 
In order to determine the timing of the strongest energy deposition, we use the derivative of the GOES SXR flux according to the so-called Neupert effect \citep{Neupert1968,VeronigEtAl2005}. Table \ref{tab:ResultsTable} lists the selected flares, together with additional information (times, position, class of the flares, associated CMEs).

All the images were rotated to solar north and were corrected for solar differential rotation. A subregion around the flare area was selected and all of the H$\alpha$ filtergrams were co-aligned with the first image of the time series using cross-correlation techniques. The MDI and HMI magnetograms were re-binned to the pixel scale of the H$\alpha$ images and were co-aligned with the first H$\alpha$ filtergram of the sequence, using the corresponding MDI or HMI continuum images. In addition, the H$\alpha$ images were normalized and filtered in order to handle large-scale intensity differences, \textit{i.e.} darkening due to clouds (for a detailed description see \citealp{PoetziEtAl2015} and \citealp{TschernitzEtAl2017}). All data were prepared and reduced using the instrument's data reduction routines in the SSW distribution.

\section{Analysis\label{DataAnalysis}}

\begin{figure}
\centerline{\includegraphics[width=0.8\textwidth,clip=]{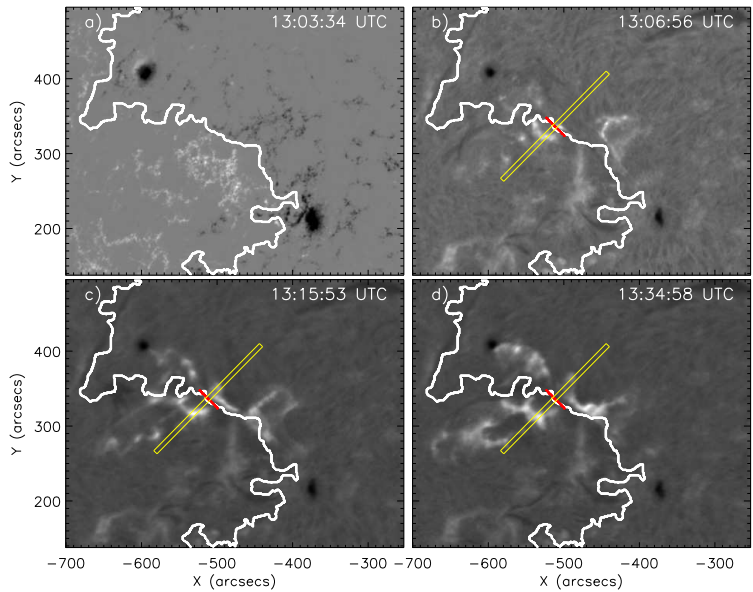}}
              \caption{M1.1/2N eruptive flare on 9 November 2011. (a) LOS magnetic field scaled to $\pm1000$\,G with the PIL indicated by the white line. (b)\,--\,(d) Three H$\alpha$ images at different times. The white line is the PIL and the red line is a linear fit of the local PIL. The yellow rectangle is perpendicular to the locally fitted PIL, indicating the direction in which the ribbons are tracked. See \textsf{Movie1.mp4} in the Electronic Supplementary Material.}
\label{20111109FlareEvolution}
\end{figure}

\begin{figure}
\centerline{\includegraphics[width=0.85\textwidth,clip=]{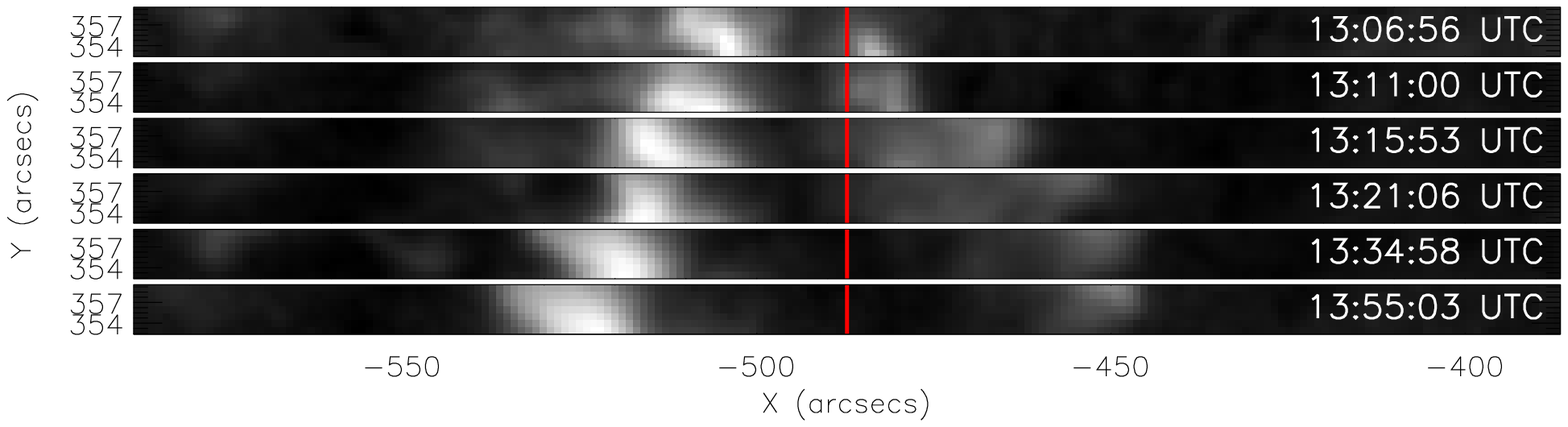}}
\centerline{\includegraphics[width=0.85\textwidth,clip=]{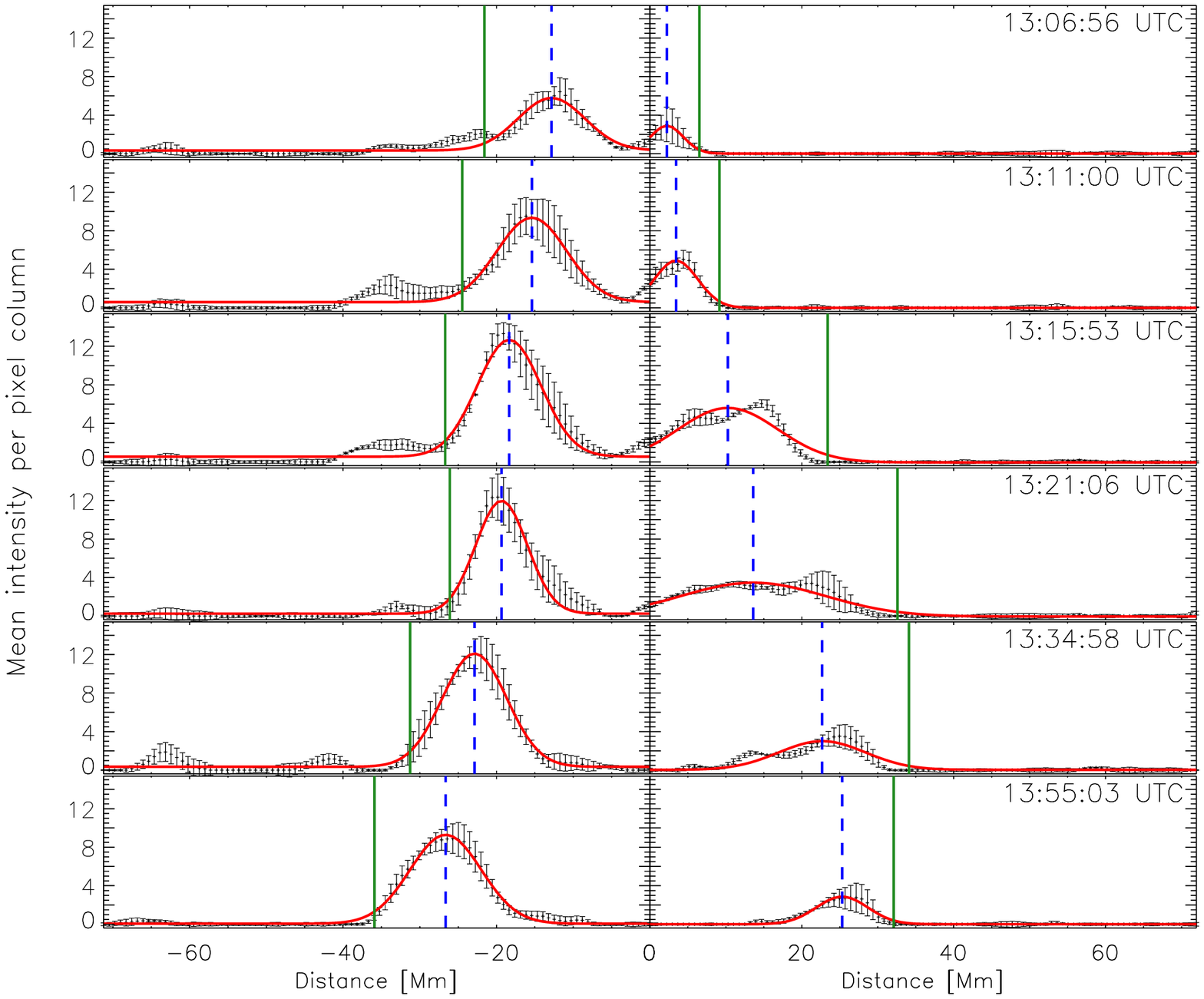}}
              \caption{Top panels: Selected subregion (\textit{cf.} yellow rectangle in Figure \ref{20111109FlareEvolution}), used to track the flare ribbons of the M1.1/2N eruptive flare on 9 November 2011, is shown for six time steps. The red-vertical line indicates the PIL. Bottom panels: Intensity profiles for the local flare ribbon on both sides from the PIL, derived from the subregion plotted at the top. Zero value indicates the position of the PIL. The black points with the error bars represent the mean intensity values and the errors are the standard deviation of the pixel intensities in one column. Red curve: Gaussian fit. Dashed vertical blue line: peak of the Gaussian fit. Vertical green line: Front of the Gaussian fit (defined as peak plus $2 \sigma$). See \textsf{Movie2.mp4} and \textsf{Movie3.mp4} in the Electronic Supplementary Material.}
\label{20111109Stripes}
\end{figure}

In the following, the method to track the flare ribbons is shown using two example flares. The first example flare is the M1.1/2N eruptive flare on 9 November 2011 (see Figure \ref{20111109FlareEvolution}). From the pre-flare HMI line-of-sight magnetic field we retrieve the flare-relevant polarity inversion line (PIL) using the IDL $\tt contour$ procedure. We then manually select a position along the PIL, from where we track the ribbon motion. Whenever possible, the slit locations are selected in such a way that both flare ribbons are well pronounced and can be tracked simultaneously in a direction perpendicular to the PIL. The white line in Figure \ref{20111109FlareEvolution} represents the PIL, the red line is a linear fit to the PIL, locally around the chosen position. The yellow rectangle indicates the subregion (length of $200''$ and width of $6''$) used to track the ribbons, perpendicular to the local PIL. The top panels in Figure \ref{20111109Stripes} show the extracted subregions within the H$\alpha$ maps for six time steps and the bottom panels show the mean intensity profiles derived along the extracted subregions. For this purpose, the mean value of each pixel column is calculated and fitted by a Gaussian (red curve). We derive the position of the leading front (vertical green line) of the ribbon by taking the peak of the Gaussian fit (dashed vertical blue line) plus $2 \sigma$. For a detailed description of the Gaussian fit function and the uncertainties see Appendix \ref{AppendixA}.

\begin{figure}
\centerline{\includegraphics[width=0.8\textwidth,clip=]{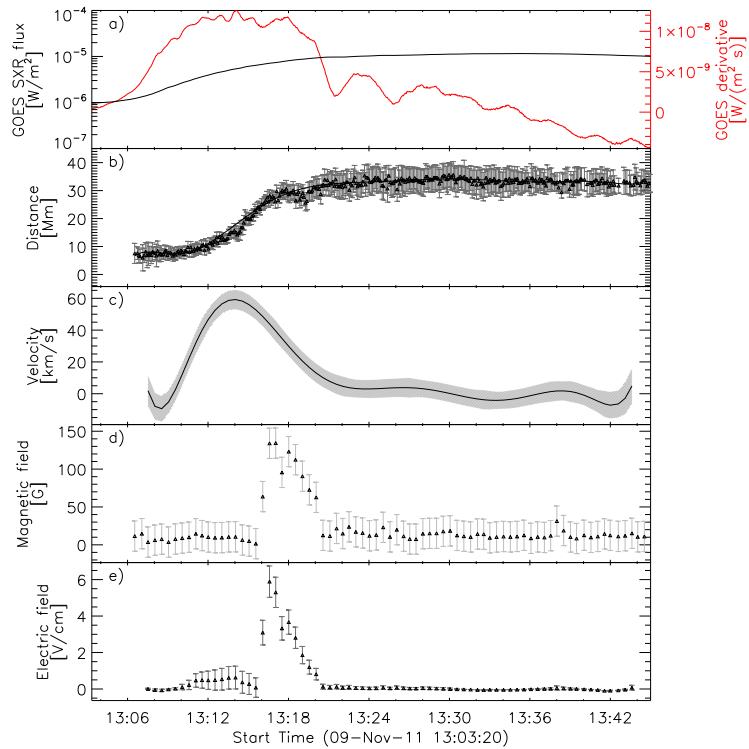}}
              \caption{Flare parameters determined for the northern (right) ribbon of the M1.1/2N eruptive flare on 9 November 2011. From top to bottom. (a) GOES $1-8$\,$\textrm{Å}$ SXR flux (black) and its derivative (red). (b) Distance of the ribbon for each time step with uncertainties and the corresponding polynomial fits. (c) Separation velocity of the ribbon with uncertainties. (d) Binned LOS magnetic-field values at the positions of the leading front with uncertainties. (e) Calculated flare electric field for the leading front with uncertainties.}
\label{20111109SummaryPlot}
\end{figure}

Figure \ref{20111109SummaryPlot} shows the summary plot of the flare ribbon detection and analysis, for the northern ribbon only (\textit{cf.} Figure \ref{20111109FlareEvolution}), \textit{i.e.}, for the intensity profiles shown on the right side of the PIL in Figure \ref{20111109Stripes}. It shows from top to bottom: a) GOES 1 \textminus{} 8\,$\textrm{Å}$ SXR flux (black) and its temporal derivative (red). b) The distance of the flare-ribbon leading front with respect to the PIL. Since we are interested in the overall ribbon motion and to improve the statistics, we binned the distance values to intervals covering 30 seconds and performed a polynomial fitting to the distance--time curve. For this particular flare we used a polynomial fit of tenth order. c) The velocities of the flare-ribbon separation for the leading front, obtained by the time derivative of the polynomial fit to the distance--time data. d) The underlying mean magnetic field at the position of the leading front of the flare ribbon, assuming an uncertainty of 20\,G. In order to account for projection effects, we apply a correction of $B_{\mathrm{n}} = B_{\mathrm{LOS}}/$cos($u$), where $u$ is the angular distance to the solar disc center and $B\textsubscript{{LOS}}$ is the line-of-sight magnetic-field strength. $u$ is calculated using the heliographic latitude and heliographic longitude listed in the KSO flare reports. e) The reconnection electric field derived using Equation \ref{eq:E=00003DvxB-1}.

\begin{figure}
\centerline{\includegraphics[width=1.0\textwidth,clip=]{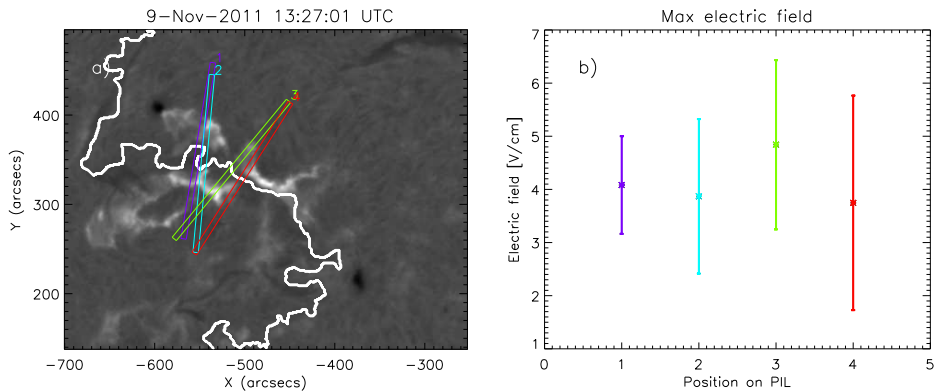}}
              \caption{(a) H$\alpha$ snapshot of the M1.1/2N eruptive flare on 9 November 2011 at the peak time. The rectangles in different colors represent different directions perpendicular to the PIL (in white) which are used to track the flare ribbons. (b) Maximum electric-field strength along different positions perpendicular to the PIL.}
\label{20111109DiffPos}
\end{figure}

This flare shows a correlation of the ribbon-separation velocity and the derivative of the GOES flux. At the time when the derivative of the GOES flux has its maximum, the ribbons are moving faster away from the PIL, reaching speeds up to 60\,km\,s$^{-1}$. With a relatively weak underlying photospheric magnetic field, which has a maximum of about 150\,G, we obtain a maximum electric field of roughly 6\,V\,cm$^{-1}$. The evolution of the electric field seems to be more affected by the magnetic field swept by the flare ribbons than by the ribbon-separation velocity (\textit{cf.} Figure \ref{20111109SummaryPlot}c\,--\,e).

The particular choice of the direction used to follow the flare ribbons on either side of the PIL may influence our results, including the minimum/maximum separation, the separation speed, and most importantly the maximum electric field. In order to assess the effect of the particular choice, we applied four different ribbon tracking directions for the M1.1/2N eruptive flare on 9 November 2011 (see Figure \ref{20111109DiffPos}a). In Figure \ref{20111109DiffPos}b the maximum electric field deduced for the individual tracking directions is shown, ranging between 4\,V\,cm$^{-1}$ and 6\,V\,cm$^{-1}$ and appearing to be quite a robust measure.

\begin{figure}
\centerline{\includegraphics[width=0.8\textwidth,clip=]{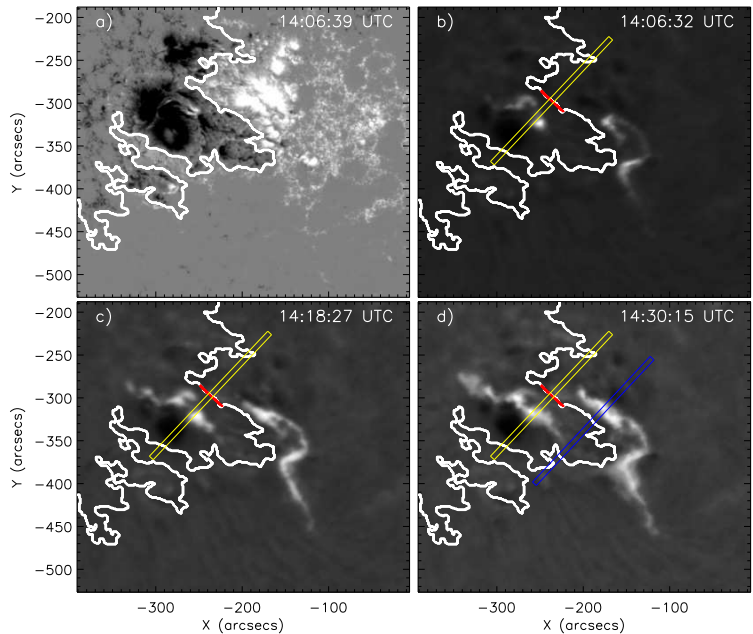}}
              \caption{X1.6/3B confined flare on 22 October 2014. (a) LOS magnetic field scaled to $\pm1000$\,G with the PIL indicated by the white line. (b)$-$(d) Three H$\alpha$ images of different times. The white line is the PIL, the red line is a linear fit of the local PIL, and the yellow rectangle is perpendicular to the locally fitted PIL, indicating the direction in which the ribbons are tracked. In panel (d) the tracking direction of the western ribbon is indicated by the blue~rectangle.}
\label{20141022FlareEvolution}
\end{figure}

\begin{figure}
\centerline{\includegraphics[width=0.85\textwidth,clip=]{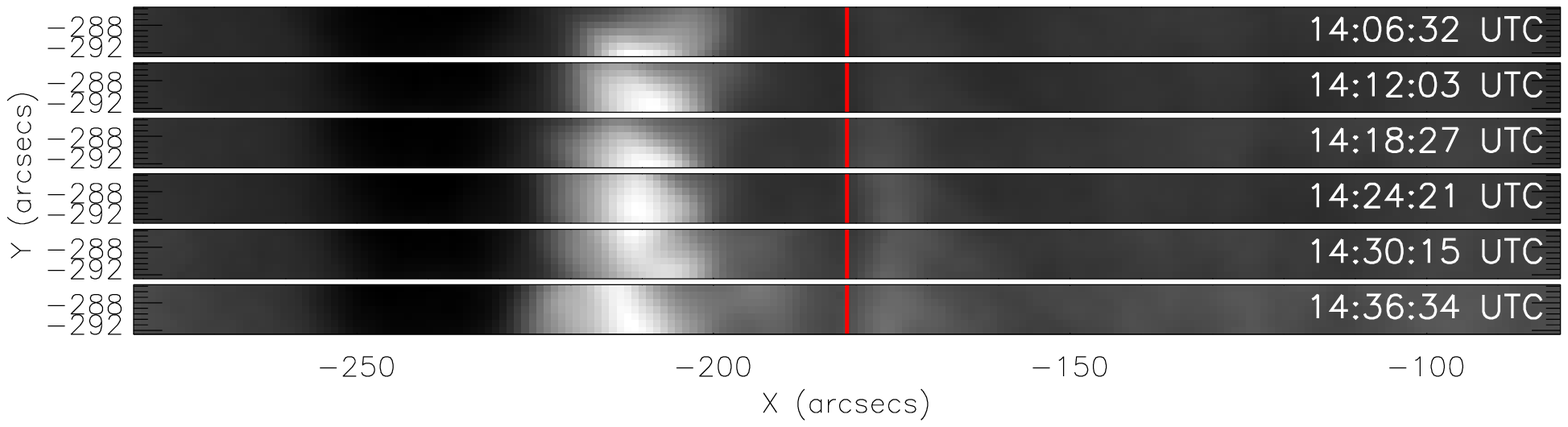}}
\centerline{\includegraphics[width=0.85\textwidth,clip=]{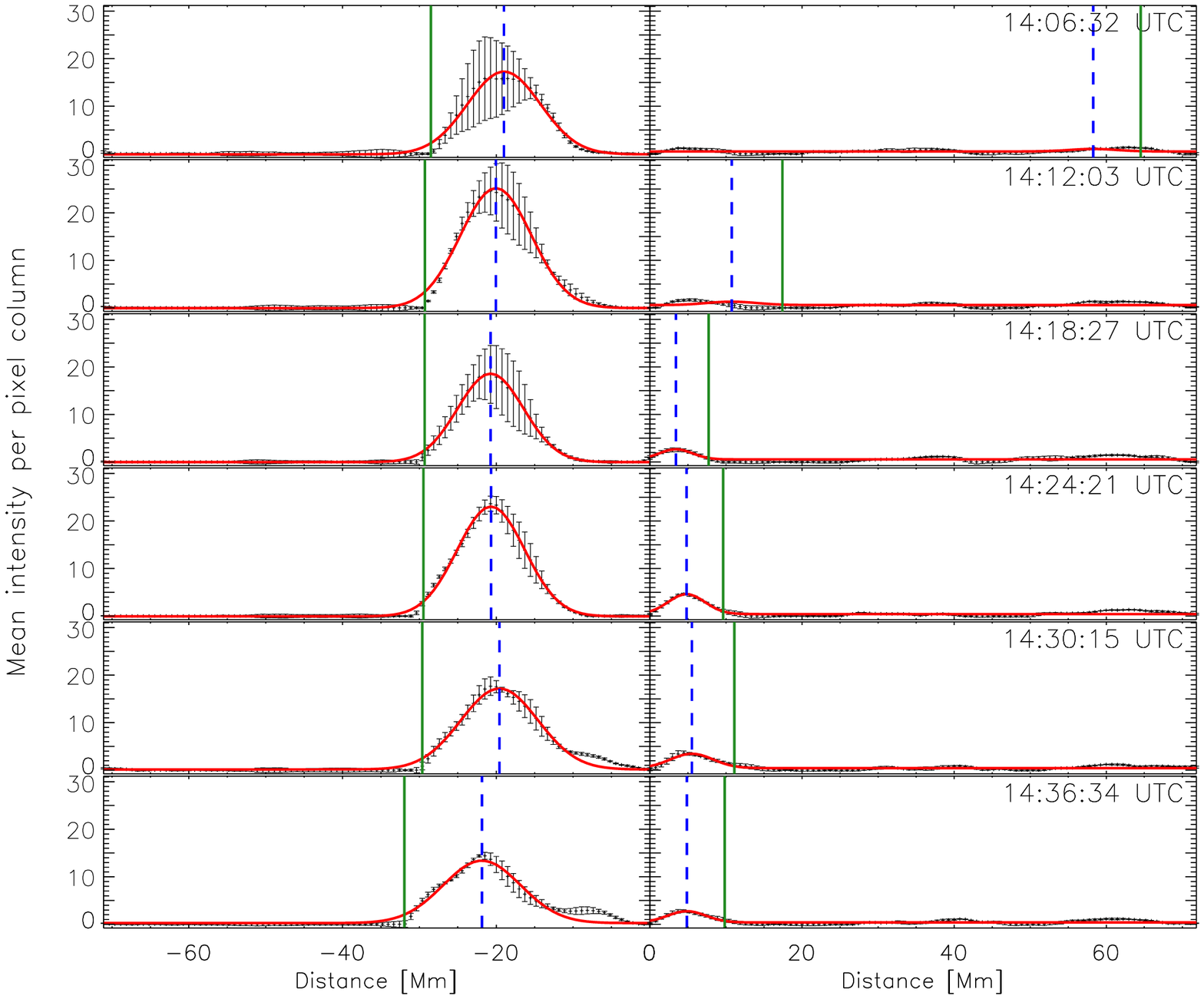}}
              \caption{Top panels: Selected subregion (\textit{cf.} yellow rectangle in Figure \ref{20141022FlareEvolution}), used to track the flare ribbons of the X1.6/3B confined flare on 22 October 2014, is shown for six time steps. The red-vertical line indicates the PIL. Bottom panels: Intensity profiles for the local flare ribbon on both sides from the PIL, derived from the subregion plotted at the top. Zero value indicates the position of the PIL. The black points with the error bars represent the mean intensity values and the errors are the standard deviation of the pixel intensities in one column. Red curve: Gaussian fit. Dashed vertical blue line: peak of the Gaussian fit. Vertical green line: Front of the Gaussian fit (defined as peak plus $2 \sigma$).} 
\label{20141022Stripes}
\end{figure}

\begin{figure}
\centerline{\includegraphics[width=0.8\textwidth,clip=]{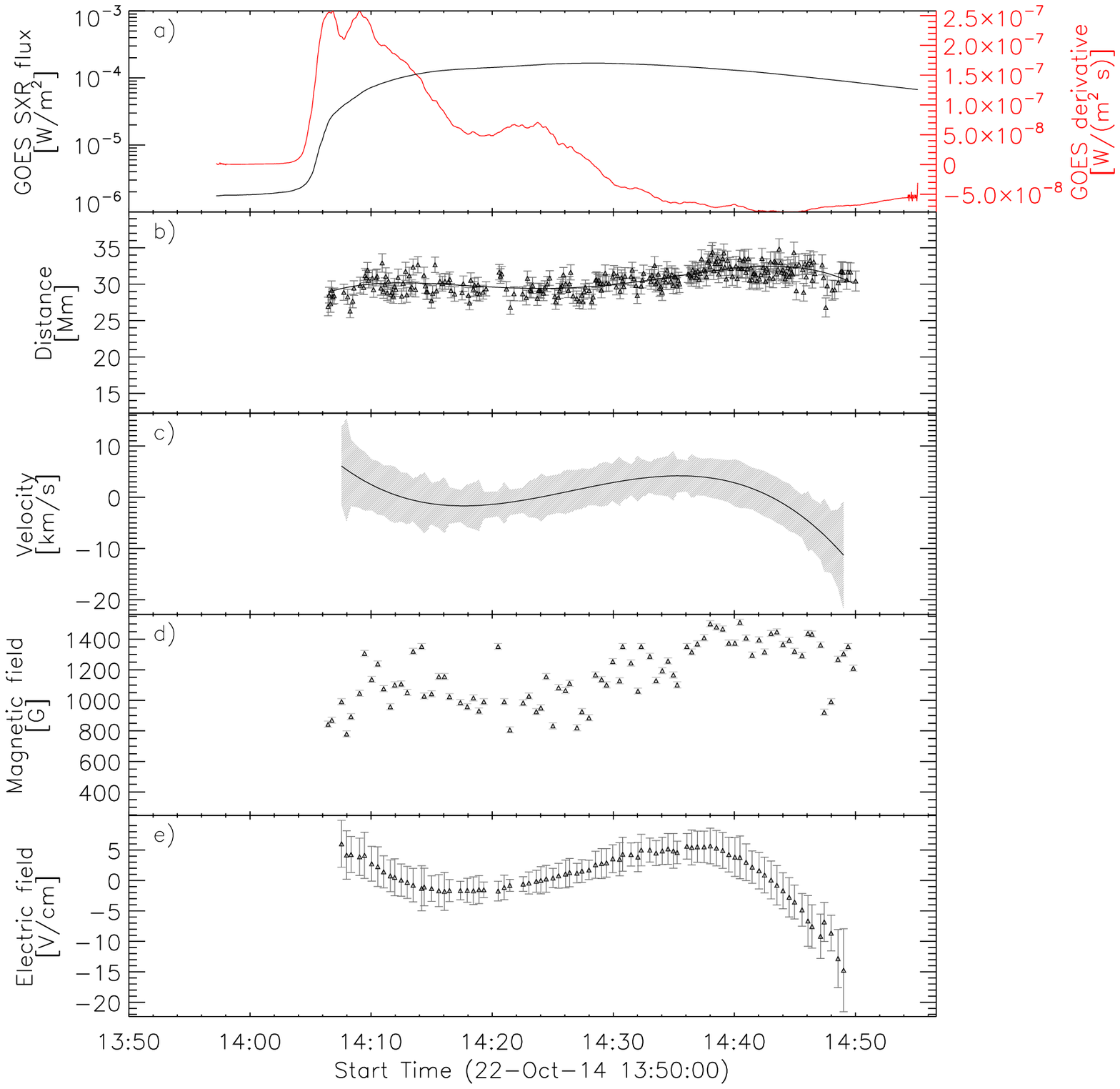}}
              \caption{Flare parameters determined for the eastern (left) ribbon of the X1.6/3B confined flare on 22 October 2014. From top to bottom. (a) GOES $1-8$\,$\textrm{Å}$ SXR flux (black) and its derivative (red). (b) Distance of the ribbon for each time step with uncertainties and the corresponding polynomial fits. (c) Separation velocity of the ribbon with uncertainties. (d) Binned LOS magnetic-field values at the positions of the leading front with uncertainties. (e) Calculated flare electric field for the leading front with uncertainties.}
\label{20141022SummaryPlot}
\end{figure}


There are events in our sample, however, for which we cannot use a single slit to follow the flare ribbons simultaneously on either side of the PIL. As an example of such a case, we show snapshots of the X1.6/3B flare on 22 October 2014 (for details see Table \ref{tab:ResultsTable}) in Figure \ref{20141022FlareEvolution}. It can be see, that the ribbons do not appear \textit{vis-à-vis} to each other, considering any direction perpendicular to the PIL, but they are strongly sheared. In such cases, we used different tracking directions for the two ribbons (see Figure \ref{20141022FlareEvolution}d). While the negative-polarity (eastern) ribbon is tracked within the subregion outlined by the yellow rectangle (see also Figure \ref{20141022Stripes}), we use the subregion outlined in blue for analysis of the positive-polarity (western) ribbon. Figure \ref{20141022SummaryPlot} shows the summary plot for the eastern ribbon of the X1.6/3B confined flare on 22 October 2014. It shows that the ribbon only marginally separates from the PIL and the peak of the separation velocity is $\approx$\,6\,km\,s$^{-1}$. However, the ribbons cover a region with a strong underlying magnetic field (up to 1600\,G; see also \citealp{VeronigPolanec2015}) and therefore, also in this event the maximum electric field reaches 6\,V\,cm$^{-1}$. \\

\section{Results}
The analysis described in Section \ref{DataAnalysis} has been performed for all flares of our event sample. For the statistical analysis we derive the minimum ribbon distance, the maximum ribbon distance, the maximum ribbon-separation velocity, the peak photospheric magnetic-field strength swept by the flare ribbons and the maximum coronal electric field. We obtained the minimum ribbon distance by summing up the minimum distance derived from the polynomial fits to the time-distance curves (\textit{cf.} Figure \ref{20111109SummaryPlot}b) for both flare ribbons. Hence, it gives an estimate of the ribbon distance at the start of the flare. The same procedure was applied for the maximum ribbon distance, but this time the maximum distance derived from the polynomial fits was summed up. The ribbon separation indicates how far the ribbons move apart from each other and is calculated by subtracting the minimum ribbon distance from the maximum ribbon distance. To get the maximum ribbon-separation speed, we compared the maximum separation velocities of both ribbons and considered only the faster ribbon (\textit{cf.} Figure \ref{20111109SummaryPlot}c). To represent a characteristic value for the underlying photospheric magnetic field, we took the magnetic-field strength at the front of the flare ribbon at the time when the coronal electric field (calculated using Equation \ref{eq:E=00003DvxB-1}) had its maximum, \textit{i.e.} at the time of the peak in Figure \ref{20111109SummaryPlot}e. Therefore, it is termed $B_{\mathrm{E}}$ in the following. Note that the product of the maximum ribbon-separation speed and the magnetic field does not necessarily result in the maximum electric field. This is because the highest ribbon-separation speeds may not necessarily occur at the same time when the ribbons are anchored in the strongest magnetic fields. The results for all the flares under study are summarized in Table \ref{tab:ResultsTable}.
 
As explained above, there are events whose flare ribbons do not appear \textit{vis-à-vis} of the PIL but are strongly sheared (see Figure \ref{20141022FlareEvolution} for an example). In such cases (indicated by an asterix in Table \ref{tab:ResultsTable}), we performed the ribbon analysis separately along two different paths (one for each polarity region). In these cases the values for minimum and maximum ribbon distance do not give the actual distance of the ribbons, but represent the sum of the individually tracked ribbons, with respect to the PIL.

\subsection{Distributions of the Flare-Ribbon Parameters}

\begin{figure}
\centerline{\includegraphics[width=1.0\textwidth,clip=]{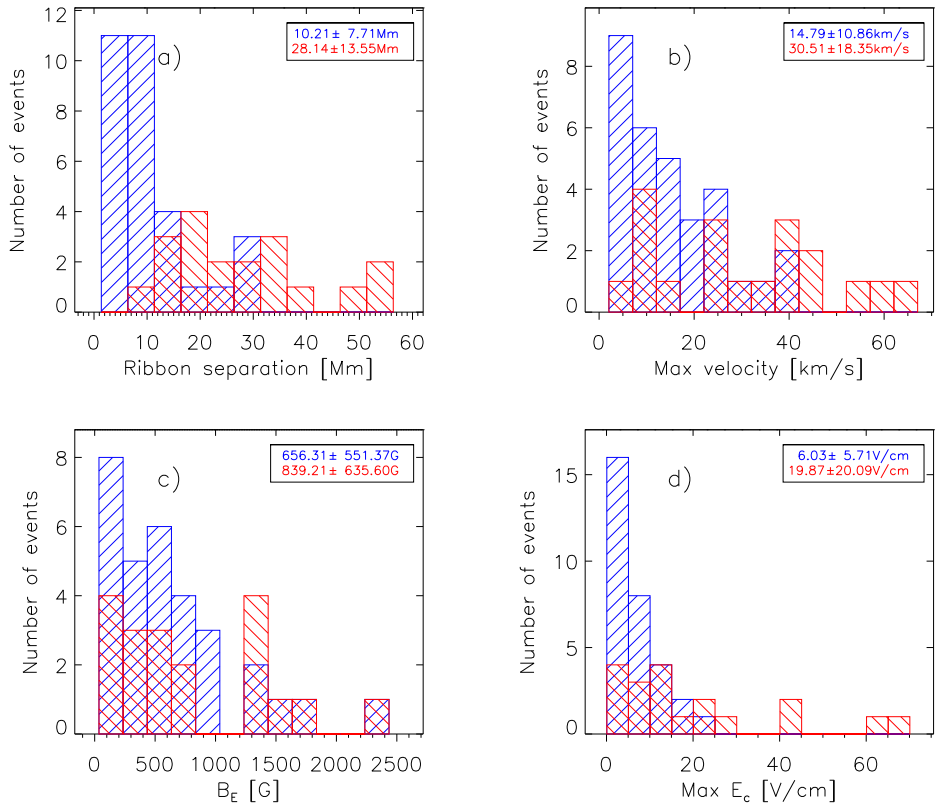}}
              \caption{Distributions of the characteristic ribbon parameters of confined (blue) and eruptive (red) flares with the mean and standard deviation in the inset. (a) Ribbon separation, (b) maximum ribbon-separation velocity, (c) magnetic-field strength [$B_{\mathrm{E}}$] at the time of the maximum electric field, (d) maximum electric field.}
\label{Histograms}
\end{figure}

Figure \ref{Histograms}a shows the distribution of the ribbon separation, indicating how far the ribbons move apart from each other during the flare. All of the eruptive flares reveal ribbon separations >\,10\,Mm. Approximately 40\,\% of eruptive flares even show a ribbon separation >\,30\,Mm. In contrast to eruptive flares, about 70\,\% of confined flares reveal a ribbon separation <\,10\,Mm; no confined flare shows a ribbon separation >\,30\,Mm. 

Figure \ref{Histograms}b presents the distribution of the maximum ribbon-separation speeds. Eruptive flares show a broad range, from 3\,km\,s$^{-1}$ up to 63\,km\,s$^{-1}$. 20\,\% of eruptive flares have maximum ribbon separation velocities >\,40\,km\,s$^{-1}$, while the separation speeds of the flare ribbons in confined events seem never to exceed $\approx$\,40\,km\,s$^{-1}$.

For 38 out of 50 flares, the strength of the photospheric magnetic field swept by the flare ribbons is <\,1000\,G (Figure \ref{Histograms}c). The distribution for confined and eruptive flares is similar, indicating that both can appear in either weak or strong magnetic fields. $B_{\mathrm{E}}$ can reach values up to almost 2500\,G (M1.2/2F confined flare on 15 January 2005). 

Figure \ref{Histograms}d shows, that roughly 50\,\% of confined flares have an electric-field strength <\,5\,V\,cm$^{-1}$, and $E_{\mathrm{c}}$ of only one confined flare exceeds 20\,V\,cm$^{-1}$. Except for four eruptive flares $E_{\mathrm{c}}$ is only found in the range less than 30\,V\,cm$^{-1}$. We find a mean electric-field strength of $6.0\pm5.7$\,V\,cm$^{-1}$ for confined flares and $19.9\pm20.1$\,V\,cm$^{-1}$ for eruptive flares. The electric-field strengths obtained in this study range from $\approx$\,0.1\,V\,cm$^{-1}$ (C1.5/SF confined flare on 29 November 2013) up to $\approx$\,70\,V\,cm$^{-1}$ for the most powerful flare under study (X17.2/4B flare on 28 October 2003), covering almost two orders of magnitude.

\subsection{Correlations of the Flare-Ribbon Parameters}

\begin{figure}
\centerline{\includegraphics[width=1.0\textwidth,clip=]{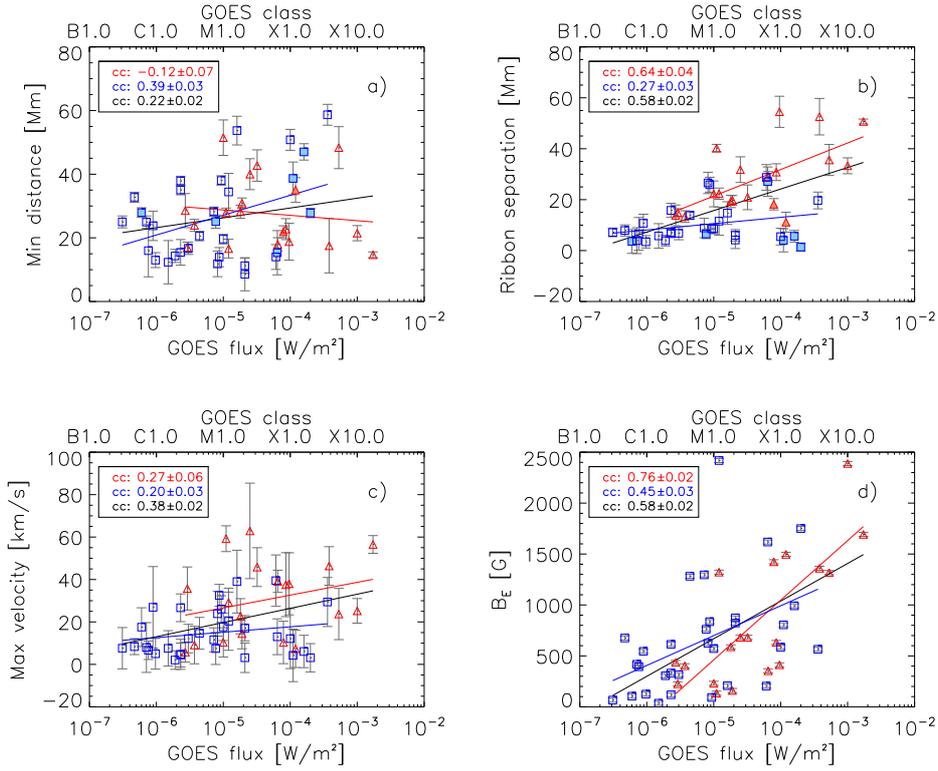}}
              \caption{Dependence of the characteristic flare-ribbon parameters as a function of the GOES class. Blue squares correspond to confined flares and red triangles to eruptive flares. The solid lines in red, blue, and black represent the linear fit of eruptive, confined, and all flares (eruptive and confined), respectively. For eight flares
(six confined, two eruptive), the analysis of each ribbon is done separately. These flares are represented by filled symbols. (a) Initial flare-ribbon distance, (b) ribbon-separation distance, (c) maximum ribbon-separation velocity, (d) magnetic-field strength [$B_{\mathrm{E}}$] at the time of the maximum electric field.}
\label{ScatterPlots}
\end{figure}

\begin{figure}
\centerline{\includegraphics[width=0.6\textwidth,clip=]{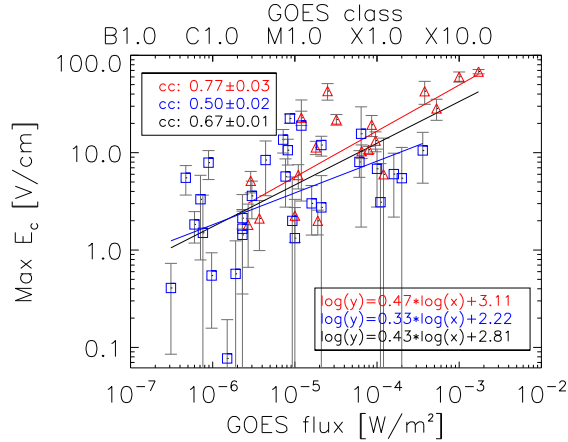}}
              \caption{Dependence of the maximum electric field on the GOES class. Blue squares correspond to confined flares and red triangles to eruptive flares. In the bottom right corner the equations for the linear fits can be found.}
\label{SCMaxE}
\end{figure}

Figures \ref{ScatterPlots} and \ref{SCMaxE} show the correlations of the characteristic flare-ribbon parameters (minimum ribbon distance, ribbon separation, maximum ribbon-separation velocity [$B_{\mathrm{E}}$] and $E_{\mathrm{c}}$), as a function of the GOES class. The solid lines represent a linear fit to the individual distributions and the corresponding correlation coefficients obtained are indicated in the left upper corner of each panel. Note that in case of a linear or log--log plot, the fit and the correlation coefficient is also calculated either in linear or log--log space. We obtain the uncertainties for the correlation coefficients using a bootstrap method. Therefore we exclude every data point once and calculate the standard deviation of all the obtained correlation coefficients. The parameters for the linear fits are listed in Table \ref{tab:FitParameters} of Appendix~\ref{AppendixB}. 

Figure \ref{ScatterPlots}a shows the minimum distance as a function of the GOES flare class. Flares ribbons that do not appear \textit{vis-à-vis} to each other are represented by filled symbols. Figure \ref{ScatterPlots}a indicates that the initial flare ribbon distance is very weakly depending on the GOES flux, \textit{i.e.}, the strength of the flare ($cc\textsubscript{{all}} = 0.22~\pm~0.02$). Weak and powerful flares can have either small or large initial ribbon distances. For eruptive flares, we find a very low correlation between the initial separation and the GOES flux ($cc\textsubscript{{eruptive}} = -0.12~\pm~0.07$), indicating that in general the former is not dependent on the latter. The correlation of the minimum distance and the GOES flux for confined flares, however, does show a trend ($cc\textsubscript{{confined}} = 0.39~\pm~0.03$).

Figure \ref{ScatterPlots}b shows how the ribbon separation depends on the GOES class. Although we find a very low correlation of the ribbon separation and the flare strength for confined events ($cc\textsubscript{{confined}} = 0.27~\pm~0.03$), the distribution is clearly separated from that of the eruptive flares. In particular, they show a lower ribbon separation for a given flare class. The ribbons of eruptive flares, on the other hand, tend to separate farther the more powerful a flare is ($cc\textsubscript{{eruptive}} = 0.64~\pm~0.04$). Considering all flares, the same trend can be found: The ribbons of more powerful flares tend to separate farther than the ribbons of weak flares ($cc\textsubscript{{all}} = 0.58~\pm~0.02$). However, it is important to note that this trend is mostly determined by that of the eruptive flares.

Figure \ref{ScatterPlots}c shows the dependence of the maximum ribbon-separation velocity on the GOES class, revealing a very weak correlation ($cc\textsubscript{{confined}} = 0.20~\pm~0.03$, $cc\textsubscript{{eruptive}} = 0.27~\pm~0.06$, $cc\textsubscript{{all}} = 0.38~\pm~0.02$). The ribbon-separation velocities of both confined and eruptive flares, show a large dispersion. Nonetheless, a constant vertical offset of about 10\,km\,s$^{-1}$ between eruptive and confined flares can be seen, indicating that the ribbons of eruptive flares tend to show higher maximum separation velocities than the ribbons of confined flares of the same class.

Figure \ref{ScatterPlots}d shows the magnetic field at the leading front of the flare ribbon at the time of the maximum electric field [$B_{\mathrm{E}}$] against the GOES class.  We obtain correlation coefficients of: $cc\textsubscript{{confined}} = 0.45~\pm~0.03$, $cc\textsubscript{{eruptive}} = 0.76~\pm~0.02$, and $cc\textsubscript{{all}} = 0.58~\pm~0.02$, indicating that more powerful flares tend to occur in stronger magnetic fields.

Figure \ref{SCMaxE} shows the dependence of $E_{\mathrm{c}}$ on the GOES flux. It illustrates that more powerful flares reveal higher electric-field strengths, which is true for both, confined and eruptive flares: $cc\textsubscript{{confined}} = 0.50~\pm~0.02$, $cc\textsubscript{{eruptive}} = 0.77~\pm~0.03$ and $cc\textsubscript{{all}} = 0.67~\pm~0.01$. The linear fits for eruptive flares are given in the form: $\log(y)=0.47 \log(x)+3.11$, for confined flares: $\log(y)=0.33 \log(x)+2.22$ and for all flares (eruptive and confined): $\log(y)=0.43 \log(x)+2.81$. 

\begin{figure}
\centerline{\includegraphics[width=1.0\textwidth,clip=]{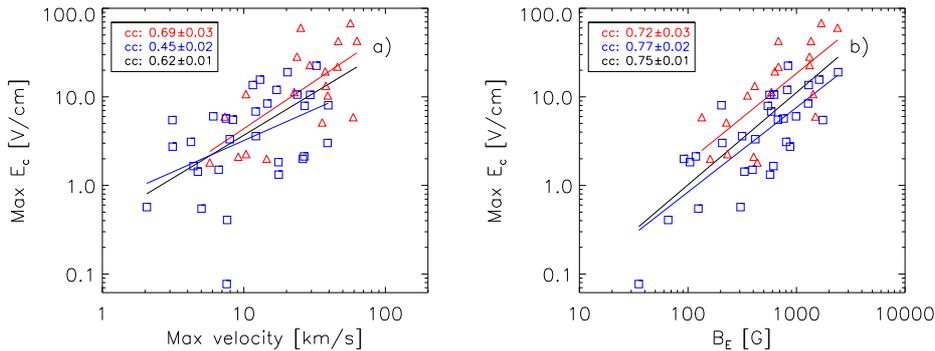}}
              \caption{Dependence of the maximum electric field on the maximum ribbon-separation velocity (left panel) and the magnetic-field strength [$B_\mathrm{E}$] at the time of the maximum electric field (right panel). Blue squares correspond to confined flares and red triangles to eruptive flares.}
\label{EvsVandB}
\end{figure}

In Figure \ref{EvsVandB} we show the correlation of $E_\mathrm{c}$ separately for the ribbon-separation speed and $B_\mathrm{E}$, in order to evaluate which of the two quantities is more strongly determining $E_\mathrm{c}$. Figure \ref{EvsVandB}a indicates that flares with higher maximum ribbon-separation speeds tend to have higher local electric-field strengths ($cc\textsubscript{{all}} = 0.62~\pm~0.01$). This is also true if considering confined and eruptive events separately ($cc\textsubscript{{confined}} = 0.45~\pm~0.02$, $cc\textsubscript{{eruptive}} = 0.69~\pm~0.03$). Considering the dependence of $E_\mathrm{c}$ on $B_\mathrm{E}$, we find higher correlations for all of the individual samples ($cc\textsubscript{{confined}} = 0.72~\pm~0.03$, $cc\textsubscript{{eruptive}} = 0.77~\pm~0.02$ and $cc\textsubscript{{all}} = 0.75~\pm~0.01$), indicating that flares occurring in regions of stronger fields tend to involve higher electric-field strengths. The constant offset in the fit curves of $E_\mathrm{c}$ against $B_\mathrm{E}$ for eruptive and confined flares can be explained by the higher ribbon-separation speeds in eruptive events.

Comparing the correlation coefficients of the reconnection electric field [$E_\mathrm{c}$] as a function of maximum ribbon-separation velocity and as a function of the magnetic field swept by the ribbons, we find that the variation of the coronal electric field is more strongly affected by differences in the involved magnetic-field strength than by the ribbon-separation speed.

\begin{figure}
\centerline{\includegraphics[width=1.0\textwidth,clip=]{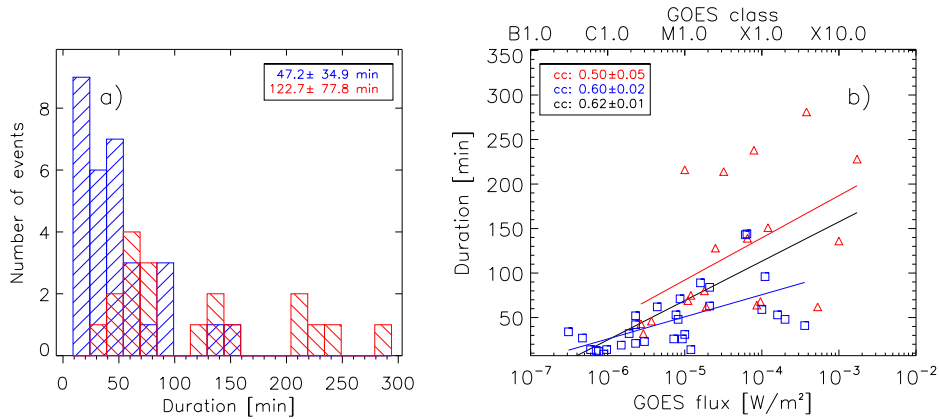}}
              \caption{(a) Distribution of the flare duration of confined (blue) and eruptive (red) flares. The mean and standard deviation are given in the inset. (b) Dependence of flare duration on the GOES class. Blue squares corresponds to confined flares and red triangles to eruptive flares. The solid lines in red, blue, and black represent the linear fit of eruptive, confined, and all flares (eruptive and confined), respectively.}
\label{DurationPlots}
\end{figure}

We also check the flare duration for significant differences between eruptive and confined flares. Figure \ref{DurationPlots}a shows the histograms of the flare duration as determined from the KSO H$\alpha$ flare reports (see KSO flare start and flare end times listed in Table \ref{tab:ResultsTable}, columns 2 and 4). We find a mean flare duration of $47.2\pm34.9$\,min for confined and $122.7\pm77.8$\,min for eruptive flares. This finding is consistent with the results of \cite{WebbHundhausen1987} who report that flares associated with CMEs tend to be of longer duration than confined flares. In Figure \ref{DurationPlots}b, we plot the flare duration as a function of GOES SXR class. This plot provides further support for this finding, as the linear fits yield a vertical offset between confined and eruptive flares of about 30\,min. However, considering solely the flare duration does not allow us to discriminate confined from eruptive flares, as the two populations of events show a significant overlap (in the range 20--100\,min; see Figure \ref{DurationPlots}a), irrespective of the flare size (compare Figure \ref{DurationPlots}b).

\begin{figure}
\centerline{\includegraphics[width=0.52\textwidth,clip=]{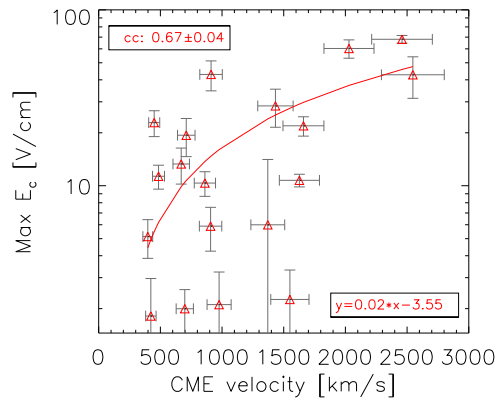}}
              \caption{Dependence of the maximum electric field on the CME velocity. In the bottom right corner the equation for the linear fit can be found.}
\label{SCMaxECMEVel}
\end{figure}

The dependence of $E_\mathrm{c}$ on the speed of the associated CME is shown in Figure~\ref{SCMaxECMEVel}, where the uncertainty of the CME velocity is assumed to be 10\,\%. We obtain a linear correlation coefficient of
$cc = 0.67~\pm~0.04$ indicating that eruptive flares with higher electric-field strengths tend to be accompanied by faster CMEs. For the linear fit we find: $E\textsubscript{{c}} = -3.55+0.02$ $V\textsubscript{{CME}}$.

\section{Summary and Discussion}

We performed a statistical study on the ribbon evolution and the coronal reconnection electric field of 50 solar flares including both confined (62\,\%) and eruptive (38\,\%) events, distributed over GOES classes B to >\,X10. We analyzed flare events that occurred from June 2000 to June 2015, homogeneously covering all H$\alpha$ and GOES flare classes. Chromospheric H$\alpha$ filtergrams from KSO, together with photospheric LOS magnetograms from MDI and HMI were used to derive the flare-ribbon separation, ribbon-separation velocity, the mean magnetic-field strength, and the reconnection electric field for the individual flare events. Our main findings are summarized as follows:
\begin{itemize}
\item Eruptive flares reveal statistically larger ribbon separation than confined flares. Almost 70\,\% of confined flares but only 10\,\% of eruptive flares show ribbon separation <\,10\,Mm. 40\,\% of eruptive flares reveal a ribbon separation >\,30\,Mm.

\item The ribbon separation of eruptive flares correlates with the GOES flux ($cc\textsubscript{{eruptive}} = 0.64$), indicating that more powerful eruptive flares separate farther. On the other hand, a very weak dependence of the ribbon separation on the GOES class for confined flares was found ($cc\textsubscript{{confined}} = 0.27$).

\item The maximum ribbon-separation velocity of eruptive flares show a wide range (up to $\approx$\,65\,km\,s$^{-1}$), whereas the majority of confined flares tend to have maximum ribbon-separation speeds <\,30\,km\,s$^{-1}$.

\item The maximum ribbon-separation velocity of both confined and eruptive flares shows almost no correlation with the GOES class ($cc\textsubscript{{confined}} = 0.20$, $cc\textsubscript{{eruptive}} = 0.27$)

\item The distribution of the maximum magnetic field swept by the flare ribbons for confined and eruptive flares is similar, indicating that both can appear in either weak or strong magnetic fields. $B_\mathrm{E}$ can reach values up to almost 2500 G.

\item For the most powerful eruptive flare under study, we find the highest coronal electric-field strengths [$E_\mathrm{c}$] up to 70\,V\,cm$^{-1}$. Only one confined flare exceeds 20\,V\,cm$^{-1}$ and except for four eruptive flares $E_\mathrm{c}$ is always <\,30\,V\,cm$^{-1}$.

\item The coronal electric field [$E_\mathrm{c}$] shows high correlation ($cc\textsubscript{{confined}} = 0.50$, $cc\textsubscript{{eruptive}} = 0.77$) with the GOES flux. Especially for confined flares, $E_\mathrm{c}$ seems to be more strongly affected by the variation in the involved magnetic field than by the ribbon-separation velocity. 

\item Eruptive flares tend to be of longer flare duration than confined flares  (see also \citealp{WebbHundhausen1987}). However, there is also a pronounced overlap in the two distributions (in particular in the range 20--100\,min).

\item Eruptive flares with higher $E_c$ tend to be accompanied by faster CMEs ($cc = 0.67$)
\end{itemize}

\citet{SuEtAl2007} studied 50 confined and eruptive flares of GOES class M and X in the time range 1998 to 2005. They found that confined flares have larger initial ribbon distances and show almost no motion perpendicular to the PIL. We find that this can not be generalized. In our extended event sample, which also includes weak confined flares, we find also small minimum distances that are comparable to that of eruptive flares. A clear difference might only exist for flares >\,M5 (see Figure \ref{ScatterPlots}a).

However, the ribbons of about 70\,\% of confined flares do not separate farther than 10 Mm, which is in good agreement with the findings of \citet{Kurokawa1989} and \citet{SuEtAl2007}. A small ribbon separation in confined flares may indicate that the reconnecting current sheet cannot move upwards. This, however, does not exclude the possibility that confined events can also show a small initial ribbon distance (\textit{cf.} Figure \ref{ScatterPlots}a), as the latter depends on the height of the reconnection region in the corona (for a recent event study see, \textit{e.g.}, \citealp{ThalmannEtAl2015}).

\citet{JingEtAl2005} studied 13 flares (mainly M and X class, 11 eruptive and 2 confined) that occurred between September 2000 and March 2004 and found a linear correlation coefficient of $cc = 0.85$ for the maximum electric field and the GOES class. We obtain a similar result for the linear correlation coefficient when considering only eruptive flares ($cc\textsubscript{{eruptive}} = 0.83~\pm~0.03$ in lin--lin space, which corresponds to $cc\textsubscript{{eruptive}} = 0.77~\pm~0.03$ in log--lin space). Considering both, confined and eruptive events, this dependence is weaker ($cc\textsubscript{{all}} = 0.67~\pm~0.01$; see Figure \ref{SCMaxE}), underlining the importance of discriminating flares in terms of their eruptivity. \citet{JingEtAl2005} also related the electric field to CME velocity. Both the linear relationship between the two parameters and the correlation coefficient of the event samples match nicely with our findings (\textit{cf.} Figure \ref{SCMaxE} and Figure 5 of \citealp{JingEtAl2005}), indicating that eruptive flares with higher $E_\mathrm{c}$ tend to be accompanied by faster CMEs.

\citet{ToriumiEtAl2017} performed a statistical study of 51 $\ge$ M5.0 flares using AIA 1600~\textrm{Å} data and found a very weak correlation ($cc = 0.20$) between the GOES peak flux and the flare ribbon distance. Even though they defined the ribbon distance in a different way (geometrical centroids of the ribbons in the two polarities), the result is comparable with our study ($cc\textsubscript{all} = 0.22$). However, we find that the GOES peak flux correlates better with the ribbon separation, \textit{i.e.} how far do the flare ribbons move apart from each other ($cc\textsubscript{all} = 0.58$), but this trend is mostly determined by eruptive events. In an accompanying paper, \citet{TschernitzEtAl2017} used the same data set as in our work to study the reconnection fluxes in eruptive and confined flares. They found a similar result to that of \citet{ToriumiEtAl2017}, that confined flares of a certain GOES class have smaller ribbon areas but larger field strengths. This is in agreement with our findings of smaller ribbon-separation speeds, leading to smaller ribbon areas in confined flares.

The X3.8/3B flare on 17 January 2005 was also analyzed by \citet{TemmerEtAl2007}. They found that the local electric-field strength is not uniform along the ribbons. They tracked the ribbons along different directions and found that the highest electric fields (up to 80\,V\,cm$^{-1}$) were obtained at flare-ribbon locations where HXR footpoints are located, and the weakest electric fields ($\approx$\,3\,V\,cm$^{-1}$) were found in regions without HXR sources. For the X3.8/3B flare on 17 January 2005 we obtain $\approx$\,40\,V\,cm$^{-1}$. Comparing the two tracking directions in \citet{TemmerEtAl2007} and in this study, we find that the ribbons were probably tracked along a direction that was associated with HXR footpoints.

The X10.0/2B flare on 29 October 2003 was studied by many authors (\citealp{XuEtAl2004,JingEtAl2005,KruckerEtAl2005,LiuWang2009,YangEtAl2011}). Table 2 in \citet{YangEtAl2011} gives a summary of the reconnection electric field for this flare. The results range from 17\,V\,cm$^{-1}$ up to 71\,V\,cm$^{-1}$, whereas the highest local electric-field strengths were obtained when tracking the location of the flare ribbons that coincide with HXR sources. Since we find $E_\mathrm{c}=60$\,V\,cm$^{-1}$ for the X10.0/2B flare on 29 October 2003, we track the H$\alpha$ flare ribbons in a region of strong energy deposition.

We found a distinct correlation between the local electric field [$E_\mathrm{c}$] in the reconnecting current sheet and the GOES soft X-ray flux for both confined and eruptive flares. These findings are suggestive of energetic particles that are accelerated by the electric field in the reconnecting current sheet \citep{Litvinenko1996}. Thus, for electrons with typical energies in the HXR range of the order of 10 to 100\,keV and with the observational determined electric fields [$E_\mathrm{c}$] from 1 up to $\approx$\,70\,keV\,cm$^{-1}$ in the reconnection region, the typical length scales for the acceleration in the current sheet are 10\,m to 10\,km, which is consistent with the findings of \citet{QiuEtAl2002}. So, a larger electric field could be responsible for higher electron acceleration in solar flares, leading to stronger emission in the X-ray regime. 

Even though we find that eruptive flares reveal a statistically larger ribbon separation and higher ribbon-separation velocities, no apparent characteristic values for eruptive or confined flares are found. This may be due to the fact that the values obtained represent local quantities, whereas the characteristics of the large-scale (global) surrounding are known to also control the eruptive behavior of flares (\textit{e.g.} the structure and strength of the confining field; for a recent statistical study see \citealp{BaumgartnerEtAl2017}).

One may also seek to find answers on the causes and consequences during CME-associated flares, \textit{e.g.}, whether a higher $E_\mathrm{c}$ necessarily leads to the expulsion of a CME or whether the flare-induced formation of a CME facilitates a larger $E_\mathrm{c}$. Regardless of the flare type (confined or eruptive), we found that $E_\mathrm{c}$ is strongly correlated with the flare size (Figure \ref{SCMaxE}). However, the distributions of $E_\mathrm{c}$ for confined and eruptive flares (Figure \ref{Histograms}d) show a significant overlap for $E_\mathrm{c}$\,<\,30\,V\,cm$^{-1}$. 

If the reconnection process in confined and eruptive events were to be distinctly different, we would expect two distinctly different populations in the $E_\mathrm{c}$ diagrams. One may attribute the fact that we do not find such differences to the fact that we employ a local reconnection rate, and that possibly existing differences might be evident on a more global scale only. But also the global peak reconnection rate determined by \citet{TschernitzEtAl2017} shows no distinction for eruptive and confined flares (see Figure 7 in \citealp{TschernitzEtAl2017}). That suggests that the electric field [$E_\mathrm{c}$] alone is not a discriminating factor for a flare to be confined or eruptive. Based on our results, we are not able to address causes and consequences within the reconnection process in eruptive events (\textit{i.e.} is a larger $E_\mathrm{c}$ a cause or a consequence of a developing CME), even more given the apparent importance of other contributing factors such as the external (confining) magnetic field surrounding the flare region, as discussed above.

\citet*{WangZhang2007} studied the magnetic properties of four confined and four eruptive X-class flares in different active regions. They found that eruptive flares usually occur at the edge of an active region (AR), whereas confined flares tend to occur near the magnetic center of an AR. They also estimated for each event the magnetic flux that penetrates a vertical plane, aligned with the polarity inversion line and extending up to 1.5 solar radii (\textit{i.e.}, the horizontal flux of the confining surrounding magnetic field). Comparing the fluxes for two height regimes ($1.0-1.1$ R$_{\astrosun}$ and $1.1-1.5$ R$_{\astrosun}$) they found that the ratio of the horizontal flux in the low corona divided by that in the high corona was significantly higher for eruptive flares. Also, the theoretical work by, \textit{e.g.}, \citet*{TorokKliem2005} indicates the importance of the magnetic field surrounding the flare region in determining whether a flare is eruptive, in particular the decay index $n$ of the magnetic field, defined as the logarithmic decay of the horizontal component of the confining magnetic field above the axis of a possibly unstable flux rope. The flux system will erupt if $n$ exceeds a critical value (\eg \citealp{KliemTorok2006,ZuccarelloEtAl2015}), implying that an overlying field that decays in strength more slowly with height may result in a flare without associated CME. This is in agreement with \citet*{SunEtAl2015}, who analyzed three active regions and found that for the flare-rich but CME-poor AR 12192, the critical value of the decay index is reached much higher in the corona than for the CME-producing active regions. 

In order to shed more light on the reconnection process of solar flares, combined measurements from spacecraft at different positions in the heliosphere would be helpful. Recently, case studies using \textit{AIA} (Atmospheric Imaging Assembly) and \textit{RHESSI} (Ramaty High Energy Solar Spectroscopic Imager) have been performed where magnetic reconnection could directly be observed (\eg \citealp{SuEtAl2013, GuoEtAl2017}). The signatures of magnetic reconnection, such as plasma inflow to the current sheet, reconnection outflows, associated energy release in form of plasma heating, and particle acceleration are best observed on the solar limb. However, in these cases we cannot measure the magnetic field, which is the crucial parameter in the physics of the events. Thus spacecraft positioned at L\tss{5} or L\tss{4} in addition to spacecraft at L\tss{1} (and ground-based observations) including magnetographs at all spacecraft may provide a big step forward in better determining the govern physical processes from the observations.

\acknowledgements{}

We thank Bhuwan Joshi from the Physical Research Laboratory (PRL) for providing the USO H$\alpha$ images and Chang Liu from the NJIT Space Weather Research Lab for the NSO H$\alpha$ data. SDO data are courtesy of NASA/SDO and the AIA and HMI science teams. This study was supported by the Austrian Science Fund (FWF): P27292-N20.

\appendix 
\section{Fitting a Gaussian Function to the Intensity Profiles} \label{AppendixA}

In order to determine the leading front of the flare ribbons, we fitted a Gaussian function to the H$\alpha$ flare ribbon intensity profiles (\textit{cf.} Figure \ref{20111109Stripes}). The Gaussian function is defined as

\begin{equation}
y(x)=A_0\exp\left[-0.5\left(\frac{x-A_1}{A_2}\right)^{2}\right],
\end{equation}
where $A_0$ is the peak value of the fit function, $A_1$ is the peak centroid and $A_2$ is the standard deviation of the Gaussian function. The peak position [$x_\mathrm{p}$] and the leading front position [$x_\mathrm{l}$] are defined to be functions of the parameters of the Gaussian fit: 
\begin{equation}
x_p = A_1,
\end{equation}

\begin{equation}
x_l = A_1+2 A_2.
\end{equation}
The uncertainty of the peak position is defined as

\begin{equation}
\Delta x_\mathrm{p}=\pm\mathrm{Max}(1\mathrm{pix}, \Delta p_1),
\end{equation}
where $\Delta p_1$ is the uncertainty on the parameter $A_1$, \textit{i.e.} the centroid position of the peak. Therefore, it is at least one pixel, assumed as the minimum error due to atmospheric and seeing conditions, but it may also be larger depending on $\Delta p_1$. The uncertainty of the leading front position is defined as

\begin{equation}
\Delta x_\mathrm{f}=\pm(\Delta x_\mathrm{p}+2\Delta p_2),
\end{equation}
where $\Delta p_2$ is the uncertainty of the parameter $A_2$, \textit{i.e.} the standard deviation of the Gaussian function. 

\section{Parameters of the linear fits} \label{AppendixB}
Table \ref{tab:FitParameters} contains the parameters of the linear fits for all correlation plots.
\begin{table}[h]

\caption{Correlation coefficients and parameters of the linear fits for confined, eruptive and all (confined and eruptive) flares for the particular scatter plots shown in Figure \ref{ScatterPlots}\,--\,\ref{SCMaxECMEVel}. Note, that some fits are applied in lin--log space, and some in log--log space.}
\setlength\tabcolsep{2pt}{\tiny \par}
\begin{tiny}
\begin{tabular}{c|c|c|c}
Figure & Confined & Eruptive & All\tabularnewline
\hline
Figure \ref{ScatterPlots}a & $cc = 0.39 \pm 0.03$ & $cc = -0.12 \pm 0.07$ & $cc = 0.22 \pm 0.02$\tabularnewline
 & $y = 6.23\log(x) + 58.29$ & $y = -1.68\log(x) + 20.38$ & $y = 3.08\log(x) + 41.70$\tabularnewline
Figure \ref{ScatterPlots}b & $cc = 0.27 \pm 0.03$ & $cc = 0.64 \pm 0.04$ & $cc = 0.58 \pm 0.02$\tabularnewline
 & $y = 2.48\log(x) + 22.96$ & $y = 10.42\log(x) + 73.52$ & $y = 8.45\log(x) + 57.99$\tabularnewline
Figure \ref{ScatterPlots}c & $cc = 0.20 \pm 0.03$ & $cc = 0.27 \pm 0.06$ & $cc = 0.38 \pm 0.02$\tabularnewline
 & $y = 2.54\log(x) + 27.88$ & $y = 6.03\log(x) + 56.75$ & $y = 6.68\log(x) + 53.11$\tabularnewline
Figure \ref{ScatterPlots}d & $cc = 0.45 \pm 0.03$ & $cc = 0.762 \pm 0.02$ & $cc = 0.58 \pm 0.02$\tabularnewline
 & $y = 292.86\log(x) + 2163.52$ & $y = 585.30\log(x) + 3387.87$ & $y = 369.29\log(x) + 2515.24$\tabularnewline
Figure \ref{SCMaxE} & $cc = 0.50 \pm 0.02$ & $cc = 0.77 \pm 0.03$ & $cc = 0.67 \pm 0.01$\tabularnewline
 & $\log(y) = 0.33\log(x) + 2.22$ & $\log(y) = 0.47\log(x) + 3.11$ & $\log(y) = 0.43\log(x) + 2.81$\tabularnewline
Figure \ref{EvsVandB}a & $cc = 0.45 \pm 0.02$  &  $cc = 0.69 \pm 0.03$  & $cc = 0.62 \pm 0.01$\tabularnewline
 & $\log(y) = 0.71\log(x) - 0.20$  & $\log(y) = 1.07\log(x) - 0.43$ & $\log(y) = 0.96\log(x) - 0.4$\tabularnewline
Figure \ref{EvsVandB}b & $cc = 0.77 \pm 0.02$ & $cc = 0.72 \pm 0.03$ & $cc = 0.75 \pm 0.01$\tabularnewline
 & $\log(y) = 0.96\log(x) - 1.98$ & $\log(y) = 1.00\log(x) - 1.73$ & $\log(y) = 1.04\log(x) - 2.08$\tabularnewline
Figure \ref{DurationPlots}b & $cc = 0.60 \pm 0.02$ & $cc = 0.50 \pm 0.05$ & $cc = 0.62 \pm 0.01$ \tabularnewline
 & $y = 24.9\log(x) + 175.1$ & $y = 47.4\log(x) + 329.3$ & $y = 44.2\log(x) + 290.1$
\tabularnewline
Figure \ref{SCMaxECMEVel} & - & $cc = 0.70 \pm 0.04$ & - \tabularnewline
 & - &  $y = 0.02 \thinspace x - 6.62$ & - \tabularnewline
\hline 
\end{tabular}
\label{tab:FitParameters}
\end{tiny}
\end{table}

\bibliographystyle{spr-mp-sola}

\end{article} 

\end{document}